\documentclass{svproc}

\usepackage{url}

\usepackage{graphicx}
\usepackage{mathtools,dsfont}
\usepackage{psfrag,bbm,graphicx}
\usepackage{algorithm,algorithmic}
\usepackage{comment,balance}
\usepackage{epstopdf}
\usepackage{epsfig}
\usepackage{multicol}
\usepackage{color}
\usepackage{multirow, cite}
\usepackage{amsfonts}
\usepackage{algorithm,algorithmic}
\graphicspath{ {images/} }
\begin{document}

\mainmatter

\title{Modeling Request Patterns in\\ VoD Services with Recommendation Systems}
\titlerunning{Request Patterns and Caching}
\author{Samarth Gupta \and Sharayu Moharir$^*$}
\institute{Department of Electrical Engineering,\\
 Indian Institute of Technology Bombay, \\Mumbai, India, 400076 \\
\email{$^*$sharayum@ee.iitb.ac.in}}

\maketitle

\begin{abstract}
Video on Demand (VoD) services like Netflix and YouTube account for ever increasing fractions of Internet traffic. It is estimated that this fraction will cross $80\%$ in the next three years. Most popular VoD services have recommendation engines which recommend videos to users based on their viewing history, thus introducing time-correlation in user requests. Understanding and modeling this time-correlation in user requests is critical for network traffic engineering. The primary goal of this work is to use empirically observed properties of user requests to model the effect of recommendation engines on request patterns in VoD services. We propose a Markovian request model to capture the time-correlation in user requests and show that our model is consistent with the observations of existing empirical studies. 

Most large-scale VoD services deliver content to users via a distributed network of servers as serving users requests via geographically co-located servers reduces latency and network bandwidth consumption. The content replication policy, i.e., determining which contents to cache on the servers is a key resource allocation problem for VoD services. Recent studies show that low start-up delay is a key Quality of Service (QoS) requirement of users of VoD services. This motivates the need to pre-fetch (fetch before contents are requested) and cache content likely to be requested in the near future. Since pre-fetching leads to an increase in the network bandwidth usage, we use our Markovian model to explore the trade-offs and feasibility of implementing recommendation based pre-fetching.

\end{abstract}

\section{Introduction}


Internet usage patterns are shifting towards content distribution and sharing with Video-on-demand (VoD) services like Netflix \cite{Netflix} and YouTube \cite{Youtube} accounting for over $50\%$ of all Internet traffic. This fraction is expected to cross $80\%$ by 2019 \cite{Cisco}. Most popular VoD services provide recommendations to users which heavily influence their viewing patterns. More specifically, recommendations lead to correlation in the videos requested by a user across time. The primary goal of this work is to model the viewing patterns of users of VoD services with recommendation engines. An accurate model of usage patterns is a crucial ingredient in the design of resource allocation algorithms which effectively manage Internet traffic and ensure high Quality of Service (QoS) to the users. 

Meeting the QoS demands of users is critical for a VoD service to retain and expand its customer base. A recent study by Akamai \cite{krishnan2013video} found that users start leaving if a video takes more than two seconds to start streaming. Moreover, for each additional second of start-up delay, the rate of abandonments increases by approximately $5.8\%$. The probability of a user returning to the VoD service within one day after watching a failed video is $8\%$ versus $11\%$ after watching a normal one. Evidently, frequent start-up delays can lead to a loss of customers, thus reducing the revenue of the VoD service.


Most large-scale VoD services serve their users via Content Delivery Networks (CDNs) which have multiple servers/caches with storage and service capabilities spread across the world. Efficient use of the available storage resources, e.g., serving user requests via geographically co-located servers can enhance the QoS for the user. More specifically, a frequent cause of start-up delay is that videos requested by users are not available on geographically co-located servers, and have to be fetched from other servers \emph{after} they are requested. The delay in start-up is caused by the large geographical/network distance between the users and servers which cache the requested content. 

The goal of reducing start-up delay motivates caching policies that are aggressive in adapting the content stored on the local servers in order to minimize the probability of delayed start-up. One possible solution is to pre-fetch (fetch before videos are requested) and cache videos that are likely to be requested in the near future \cite{khemmarat2012watching, krishnappa2011feasibility, du2015analysis, plecsca2016multimedia, liang2015integrated}. Since pre-fetching leads to an increase in the bandwidth consumption of the CDN, there is a trade-off between bandwidth usage of the network and the quality of service provided to the users. We explore this trade-off in this work.


\subsection{Contributions}
The contributions of this work can be summarized as follows.  

\emph{Modeling the request process:} In a preliminary version of this work \cite{gupta2017}, we propose a Markovian model which captures the time-correlation in user requests in VoD services due to the presence of recommendation systems. We show that our model is consistent with empirically observed properties of request patterns in such VoD services \cite{cheng2007understanding, cheng2009nettube, zhou2010impact, krishnappa2015cache}. A limitation of this model is that it imposes the constraint that the recommendations are symmetric (i.e., Video A recommends Video B implies Video B recommends Video A). In this work, we generalize our model to allow for non-symmetric recommendation relationships. 

\emph{Performance evaluation of caching policies:} We study a caching policy which pre-fetches videos likely to be requested in the future in order to minimize the chance of delayed start-up. More specifically, while a user is watching a video, our policy pre-fetches a pre-determined number of the corresponding recommended videos to the local cache, thus reducing the probability that the next request from this user experiences any start-up delay. 

As discussed above, pre-fetching content reduces start-up delay, but, leads to increased bandwidth consumption. Via simulations, we explore this trade-off as a function of the relative costs of bandwidth consumption and delayed start-up. Our results characterize when pre-fetching content can lead to a reduction in the overall cost of service, even with the increased bandwidth usage.

\subsection{Organization}

The rest of the paper is organized as follows. In Section \ref{section:observed_properties}, we discuss existing literature on empirical studies of viewing patterns in VoDs with recommendation systems. In Section \ref{section:our_model}, we define our Markovian request model and discuss its properties. We describe our CDN setting in Section \ref{section:CDN} and discuss the proposed caching scheme in Section \ref{section:CachingPolicies}. In Section \ref{section:simulate}, we evaluate the performance of the proposed policy via simulations. We present our conclusions in Section \ref{section:conclusions}.
\section{Literature Review}
\label{section:observed_properties}



\subsection{Request Patterns in VoDs with Recommendations}
\label{subsection:empirical_studies}

We first summarize the observations of empirical studies which study the effect of recommendation systems on the users' viewing patterns \cite{cheng2007understanding, cheng2009nettube, zhou2010impact, krishnappa2015cache}. These studies have been conducted either by crawling the YouTube webpage \cite{zhou2010impact}, or via the Youtube API \cite{zhou2010impact}, or by collecting browsing data from university networks \cite{zhou2010impact, krishnappa2015cache}. The studies represent the relationship between videos using a directed graph, where nodes represent videos and each node has a directed edge to all the corresponding recommended videos. They focus on the properties of the graph \cite{cheng2007understanding}, the effect of the placement/rank of a video in the recommendation list of another video \cite{zhou2010impact,krishnappa2015cache}, and the effect of recommendations on the overall video popularity profile \cite{cheng2009nettube}.

\subsubsection{Small-World Recommendation Graph}
The key insight obtained in \cite{cheng2007understanding} is that the graph representing the YouTube recommendation network is small-world. We use the following definitions to formally define small-world networks.

\begin{itemize}
\item[(i)] \emph{Characteristic Path Length:} The characteristic path length of a network is defined as the mean distance between two nodes, averaged over all pairs of nodes.

\item[(ii)] \emph{Clustering Coefficient:} The clustering coefficient of a network is defined as the average fraction of pairs of neighbors of a node that are also neighbors of each other.

\end{itemize}
Small-world networks are a class of networks that are highly clustered (high clustering coefficient), like regular lattices, yet have small characteristic path lengths, like random graphs \cite{watts1998collective}. Compared to random graphs with the same average degree, small-world networks are characterized by high clustering coefficients and similar path lengths. In \cite{cheng2007understanding}, the authors use these two characteristics to conclude that the YouTube recommendation graph is small-world.

%


\subsubsection{Content Popularity Profiles}

It has been observed that content popularity for VoD services \emph{without} recommendation systems is heavy-tailed and can often be well-fitted with the Zipf distribution defined as follows: the popularity of the $i^{th}$ most popular video is proportional to $i^{-\beta}$, where $\beta$ is a positive constant called the Zipf's parameter. Typical values of $\beta$ for VoD services lie between 0.6 and 2 \cite{adamic2002zipf}.

Empirical studies have concluded that content popularity  for VoD services \emph{with} recommendation systems, e.g., YouTube, can be well-fitted with the Zipf distribution for the popular videos and popularity for the less popular videos decreases faster than the rate predicted by the Zipf distribution \cite{cheng2009nettube}.

%

\subsubsection{Click Through Rate}
The Click Through Rate (CTR) for position $r$ in the recommendation list of Video $i$ is defined as the fraction of times a user requests the video in position $r$ in the recommendation list of Video $i$ right after watching Video $i$. In \cite{zhou2010impact}, the authors found that the mean of the CTR follows the Zipf distribution as a function of $r$. In addition, Figure 3 in \cite{krishnappa2015cache} shows that the CDF of the CTR is concave. 

\subsubsection{Chain Count}
Chain count is defined as the average number of consecutive videos a user requests by clicking on videos in the recommendation list before requesting a video which is not the list of recommended videos for the video currently being watched. For YouTube, the chain count is estimated to be between 1.3 and 2.4 in \cite{krishnappa2015cache}.

\subsubsection{Degree Distribution} 
The degree distribution of the recommendation graph has been found to follow the power law. More specifically, the number of nodes with degree $k$ is approximately proportional to $k^{-3}$ \cite{sweetyshubham2016}.

\subsection{Pre-fetching based caching schemes}

Caching schemes which use pre-fetching have been shown to be beneficial for TV-on-demand and VoD services \cite{khemmarat2012watching, krishnappa2011feasibility, du2015analysis, plecsca2016multimedia, liang2015integrated}. To the best of our knowledge, none of the existing works have attempted to model the request arrival process for VoD services with recommendation systems, and instead, use trace data to evaluate the performance of the proposed policies. In addition, another key difference between the existing literature and this work is that we study the trade-off between bandwidth usage and quality of service, while most of the existing works (except \cite{du2015analysis}) focus only on the improvement in quality of service (cache hit-ratio) by pre-fetching content.   

In \cite{khemmarat2012watching}, the authors use trace data from a campus network gateway to analyze the performance of pre-fetching content to serve  YouTube requests. A key observation in \cite{khemmarat2012watching} that it is not necessary to pre-fetch complete videos to avoid start-up delays. Fetching a fraction of the video is often sufficient as the rest of the video can be fetched while the users' watch the initial part of the video. In \cite{krishnappa2011feasibility}, the authors compare the performance of pre-fetching+caching and the Least Recent Used (LRU) caching scheme which does not pre-fetch content, for Hulu (a VoD service) on a university network. 
In \cite{du2015analysis}, trace data from a Swedish TV service provider is used to evaluate the benefits of pre-fetching episodes of shows that a specific user is watching in order to reduce latency. 
In \cite{plecsca2016multimedia}, the authors study the setting where the requests arrive according to a known Markov process. They propose an MDP based pre-fetching scheme and prove its optimality. Although our work also assumes that the underlying request process is Markovian, unlike \cite{plecsca2016multimedia}, our caching policy works without the knowledge of the transition probabilities. This is an important distinction, since for VoD services like YouTube with massive content catalogs, content popularity is often time-varying and unknown \cite{SGSS14}. In \cite{liang2015integrated}, the authors study a pre-fetching and caching scheme for HTTP-based adaptive video streaming. They propose a pre-fetching and caching scheme to maximize the cache hit-ratio assuming the bandwidth between the local cache and the central server is limited. 

\section{Our Request Model}
\label{section:our_model}

In this section, we discuss our model for the request process for VoD services with recommendation systems. 

\subsection{Model Definition}
\label{subsection:definition}
We construct a directed graph $G(V,E)$, where the set $V$ consists of all the videos offered by the VoD service and an edge $e = \{i,j \} \in E$ implies that Video $j$ is one of the recommended videos for Video $i$. We then assign weights to edges. Each user's request process is a random walk on this weighted graph and therefore, the request arrival process is Markovian and can be completely described by a transition probability matrix.

We use a subset of the properties discussed in Section \ref{subsection:empirical_studies} to construct this matrix and verify that the remaining properties discussed in Section \ref{subsection:empirical_studies} are satisfied by our Markovian model. 

Motivated by the fact the empirical studies like \cite{cheng2007understanding} have found that this graph is small-world, and the degree distribution follows the power law \cite{sweetyshubham2016} we use the Barabasi-Albert model \cite{albert2002statistical} to generate a random small-world graph. Refer to Figure \ref{fig:BA} for a formal definition of the Barabasi-Albert model.

\begin{figure}[h]
	\hrule
	\vspace{0.1in}
	\begin{algorithmic}[1]
		\STATE Initialize: Generate a connected graph of $m$ nodes ($v_1$, $v_2$, ..., $v_m$). Let $v=m+1$.
		\STATE Introduce a new node $n_v$ which connects to $m$ existing nodes. These $m$ edges from $n_v$ are added in an sequential manner as follows. The probability that each of the $m$ edges from the new node go to an existing node $n_i$ is given by $p_i$ such that
		$$p_i = \dfrac{K_i}{\sum_{j} K_j}, $$ 
		where $K_i$ is the current degree of node $n_i$.
		\STATE $v = v+1$. If $v < n$, goto Step 2.
	\end{algorithmic}
	\vspace{0.1in}
	\hrule
	\caption{Barabasi-Albert Model -- \sl An algorithm to generate a random small-world graph with a degree distribution following the power law.}
	\label{fig:BA}
\end{figure}

Since the Barabasi-Albert model generates a undirected graph, we replace each edge by two directed edges to obtain a directed graph on the set of videos. This means that if $v_i$ recommends $v_j$, our model assumes that $v_j$ also recommends $v_i$. This is motivated by the fact that YouTube uses the relatedness score \cite{davidson2010youtube} for each pair of videos to determine homepage recommendations. The relatedness score of two videos is proportional to the number of times two videos are co-watched in a session. Therefore, by definition, if $v_i$ is closely related to $v_j$, $v_j$ is closely related to $v_i$.

Users can request videos via multiple sources. We divide them into two categories:
\begin{itemize}
\item[--] The first set of requests come via the recommendations made by VoD service when the user is watching a video. We introduce a quantity $P_{cont}$, defined as the probability that a user requests a recommended video after he/she finishes watching the current video. Formally, after watching a video, each user requests one of the recommended videos with probability $P_{cont}$ independent of all previous requests. By definition, the expected chain count (defined in Section \ref{subsection:empirical_studies}) is given by $1/(1-P_{cont})$. The value of $P_{cont}$ should be between 0.2 and 0.7 to be consistent with the chain count values observed in \cite{krishnappa2015cache}.

\item[--] The second set of requests come from all other sources on the Internet including the VoD homepage, the user's social networking page, etc. To model the second type of requests, we add a dummy node $n_0$ to the graph $G$. This dummy node represents all other sources of requests and is connected to all other nodes in the $G$ via two directed edges.
\end{itemize}

The next step is to assign transition probabilities corresponding to each edge in this directed graph $G(V,E)$. Let $P_{i,j}$ be the probability a node makes the transition from node $n_i$ to node $n_j$. 
\begin{itemize}
	\item[--] By definition, $P_{i,j}=0$ if $\{i,j\} \notin E$. 
	\item[--] Recall that $P_{cont}$ is the probability that a user requests one of the recommended videos after watching the current video. If not, we assume that the user goes to node $n_0$ which represents all other sources of video requests. Therefore, by definition, $P_{i,0}=1-P_{cont},$ $ \forall i>0$. 
	\item[--] Motivated by the fact that for VoD services without recommendations, content popularity follows the Zipf distribution (as discussed in Section \ref{subsection:empirical_studies}), we set the value of $P_{0,j} \propto j^{-\beta}$ for a positive constant $\beta$ called the Zipf parameter. Typical values of $\beta$  for VoD services lie between 0.6 and 2 \cite{adamic2002zipf}.
	\item[--] 
	To assign transition probabilities to edges between a video and its recommended videos, we use the distance between two videos as a measure of similarity in the content of the two videos. For each ${i,j} \in E$, $P_{i,j} \propto P_{cont}.(D(i,j))^{-{\kappa}}$, where $D(i,j) = |i-j|$ and $\kappa$ is a positive constant. We use the $P_{i,j}$s to determine the order in which the recommended videos are presented to the user. For Video $i$, we assume that the recommended videos are ordered in decreasing order $P_{i,j}$s. 
	\end{itemize}
\begin{remark}
	Our model is characterized by five parameters, namely, the total number of videos $n$, the size of the graph used in the first step of the Albert-Barabasi model (Figure \ref{fig:BA}) $m$, the Zipf parameter $\beta$, the probability that a user requests one of the recommended videos after watching the current video $P_{cont}$, and $\kappa$. 
\end{remark}

\begin{remark}
	Another way to assign transition probabilities from a video to its recommended videos is to pick a permutation of the set of recommended videos and assign transition probabilities according to the Zipf law. By construction, this Markov chain will satisfy the property that the mean CTR follows the Zipf distribution and therefore will be consistent with properties observed in Section \ref{subsection:empirical_studies}. 
	Unlike the model we propose, in this construction, the probability of requesting the $i^{th}$ ranked recommendation is the same across all videos. 
	\end{remark}

\subsection{Properties}
Our model uses the empirically observed properties that the recommendation graph is small-world, its degree distribution follows the power law, content popularity in the absence of recommendations follows the Zipf distribution, and the chain count is between 1.3 and 2.4. In this section, we verify that our Markovian model satisfies the remaining properties discussed in Section \ref{subsection:empirical_studies}.

\subsubsection{Content Popularity Profile}
\begin{figure}[t]
\centering
\includegraphics[width = 0.7\textwidth]{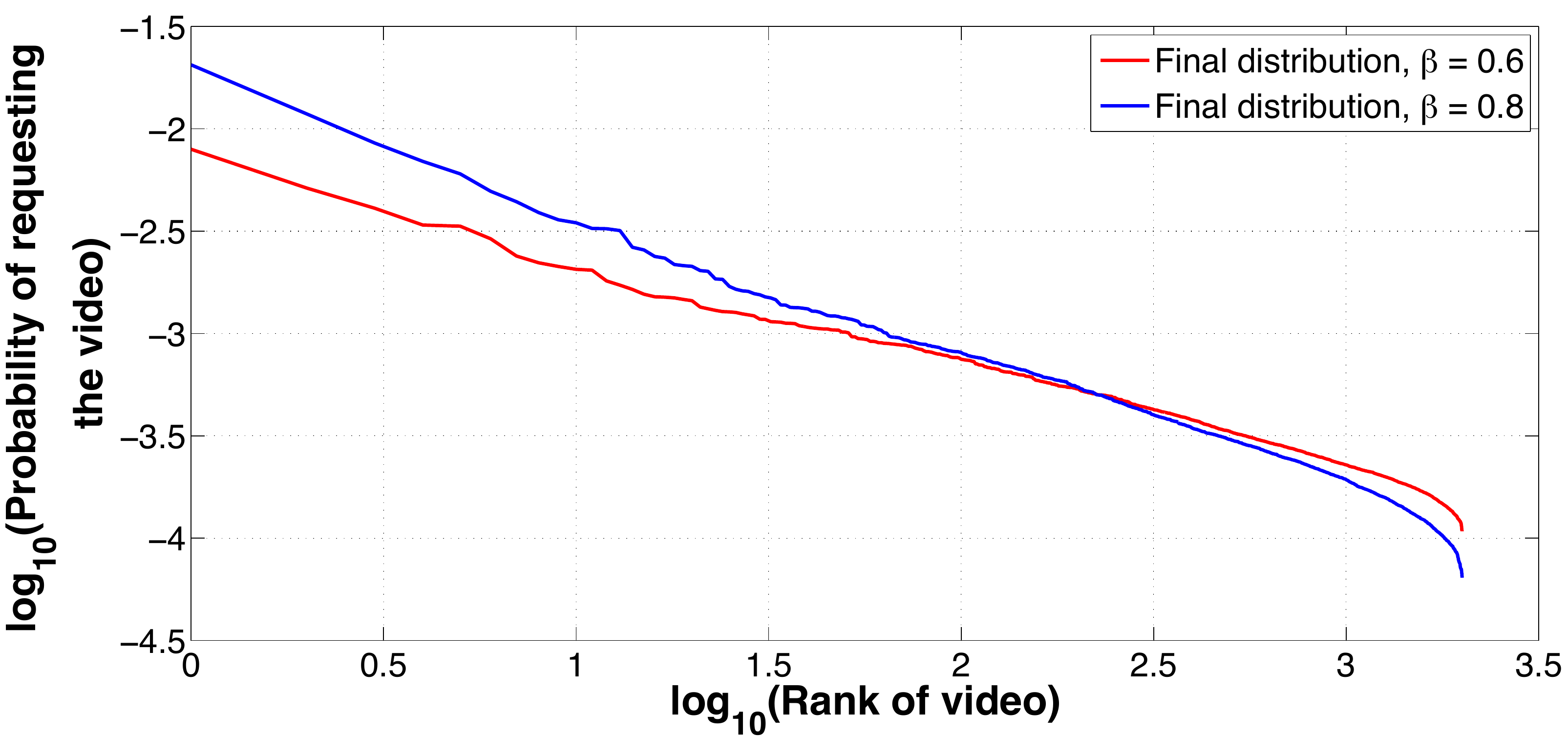}
\caption{Content popularity profile for our model with $m = 20$, $P_{cont} = 0.4$, \newline Number of videos ($n$) = 2000 and $\kappa = 0.8$. } \label{fig:final_popularity} 
\end{figure}

\begin{figure}[h]
\centering
\includegraphics[width = 0.7\textwidth]{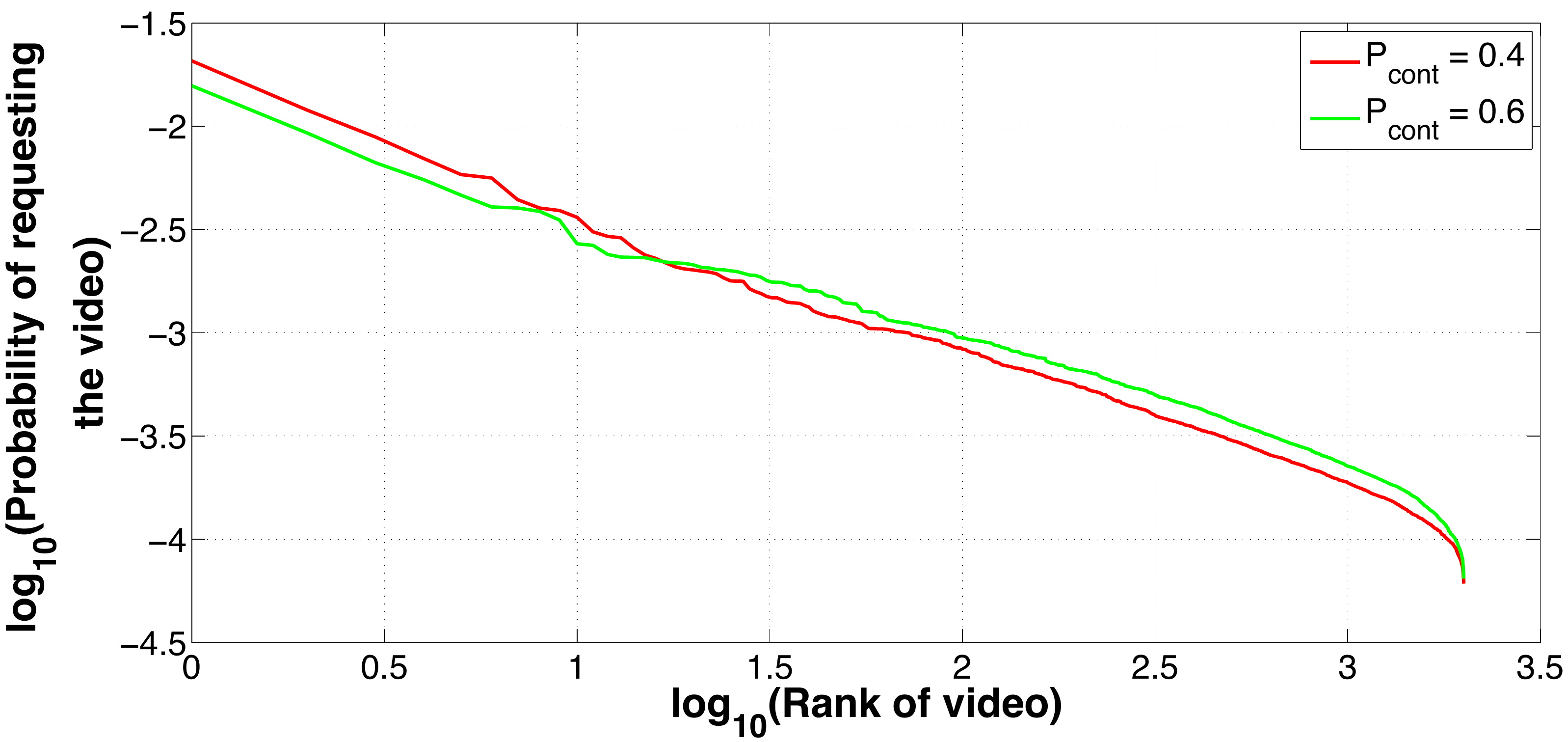}
\caption{Content popularity profile for our model with $m = 20$, Zipf parameter $(\beta) = 0.8$, Number of videos ($n$) = 2000 and $\kappa = 0.8$.} \label{fig:finalvsPcont} 
\end{figure}

The popularity of a video is the fraction of total requests for the video. Since the requests are generated by a finite state irreducible Discrete Time Markov Chain (DTMC), this is equal to the steady state probability of requesting the video. We therefore compute the content popularity profile of our model by calculating the stationary distribution of the Markov Chain. Figure \ref{fig:final_popularity} illustrates the content popularity profile for a system consisting of 2000 videos as a function of the Zipf Parameter $\beta$. Figure \ref{fig:finalvsPcont} shows how final distribution varies with $P_{cont}$.


We see that, as desired, the content popularity profile follows the Zipf distribution for the popular videos and decreases faster than as predicted by the Zipf distribution for the unpopular videos. We thus conclude that the content popularity profile for our model is consistent with the observations in \cite{cheng2009nettube}.
 
\subsubsection{Click Through Rate} 
\begin{figure}[h]
\centering
\includegraphics[width = 0.7\textwidth]{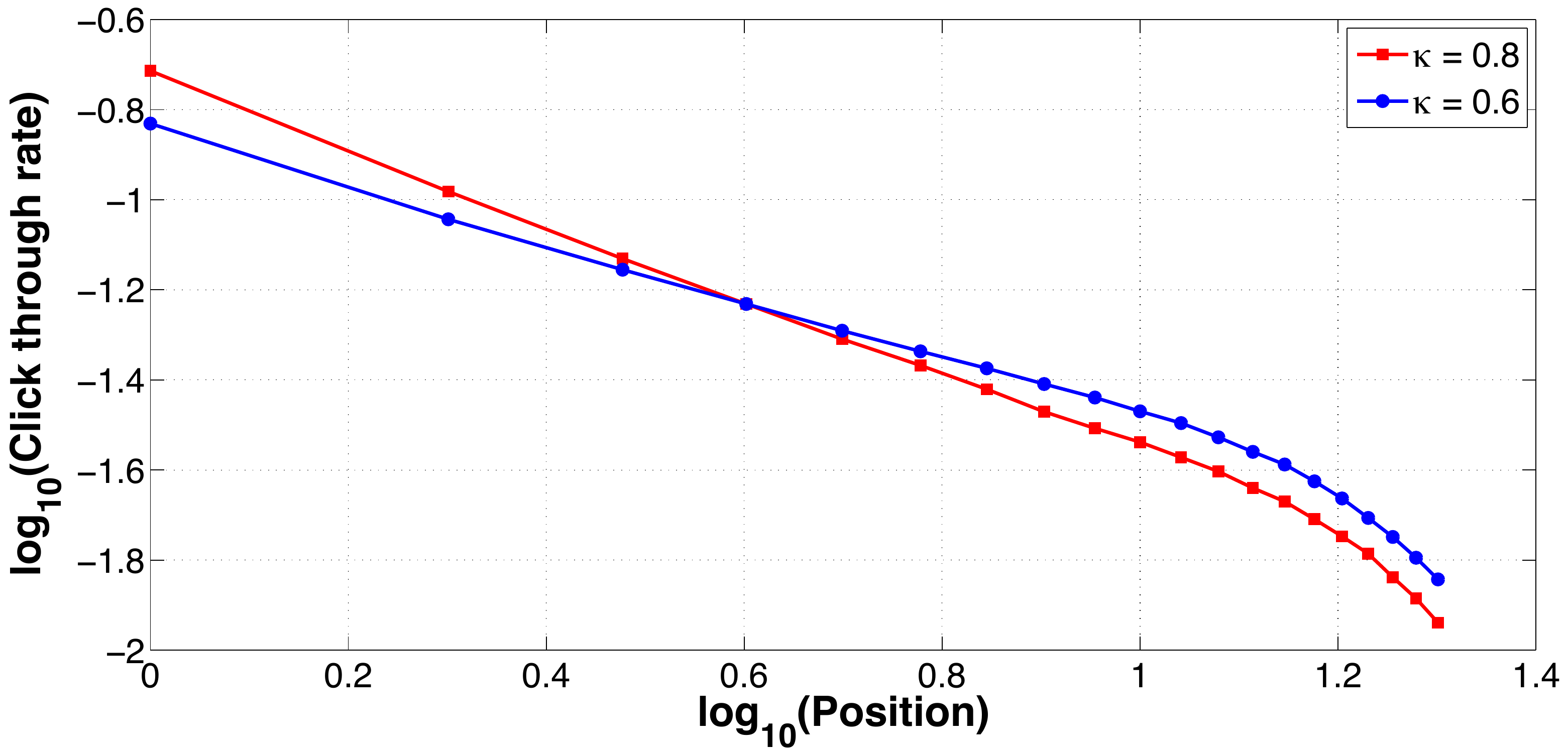}

\caption{Median Click through rate (CTR) as a function of position of video in the recommendation list for Number of videos ($n$) = 2000, $m = 20$, Zipf Parameter $\beta = 0.8$, and $P_{cont} = 0.4$. } \label{fig:median_CTR} 
\end{figure}

As discussed in Section \ref{subsection:empirical_studies}, the median Click Through Rate (CTR) follows the Zipf distribution. To verify this for our model, we compute the probability of requesting the $r^{th}$ ranked recommended video for each video. We plot the median of this quantity across all videos as a function of $r$ in Figure \ref{fig:median_CTR}. We see that the median CTR can be approximated by the Zipf distribution. Our model is therefore consistent with the observations in \cite{zhou2010impact}. Varying $\kappa$ allows us to change the slope of median CTR.

\subsubsection{CDF of Click Through Rate} 
As mentioned in Section \ref{subsection:empirical_studies}, in \cite{krishnappa2015cache}, the authors compute the Cumulative Distribution Function (CDF) of the Click Through Rate (CTR). To evaluate the CDF, we compute the CTR for the $r^{\text{th}}$ ranked video in the recommendation list as follows:
$$\text{CTR}(r) =  \displaystyle \sum_{i=1}^{n} \pi(i) \times P_{i, r^{\text{th}} \text{ ranked recommended video}}.$$

\begin{figure}[h]
\centering
\includegraphics[width = 0.7\textwidth]{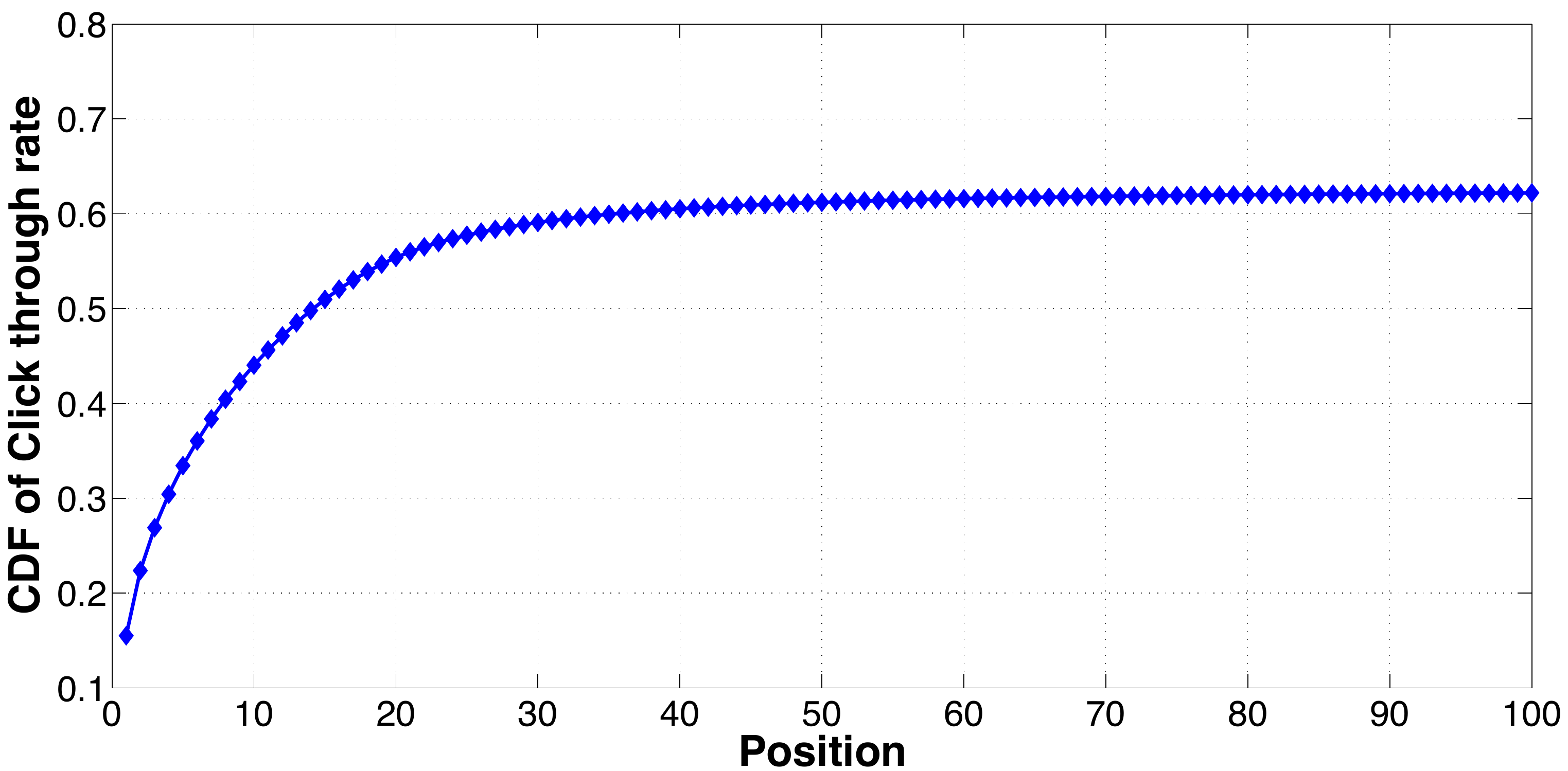}

\caption{CDF of CTR vs position in recommendation list for Number of videos ($n$) = 2000, $m = 20$, Zipf parameter $\beta = 0.8$, $\kappa = 0.8$, and $P_{cont} = 0.4$.}\label{fig:CDF_CTR_our_model} 
\end{figure}

We plot the CDF of the CTR as a function of the position $r$ in Figure \ref{fig:CDF_CTR_our_model}. Qualitatively, Figure \ref{fig:CDF_CTR_our_model} shows the same trend as observed in Figure 3 in \cite{krishnappa2015cache}.

\section{CDN Setting}
\label{section:CDN}

\begin{figure}[h]
\centering
\includegraphics[width = 0.7\textwidth]{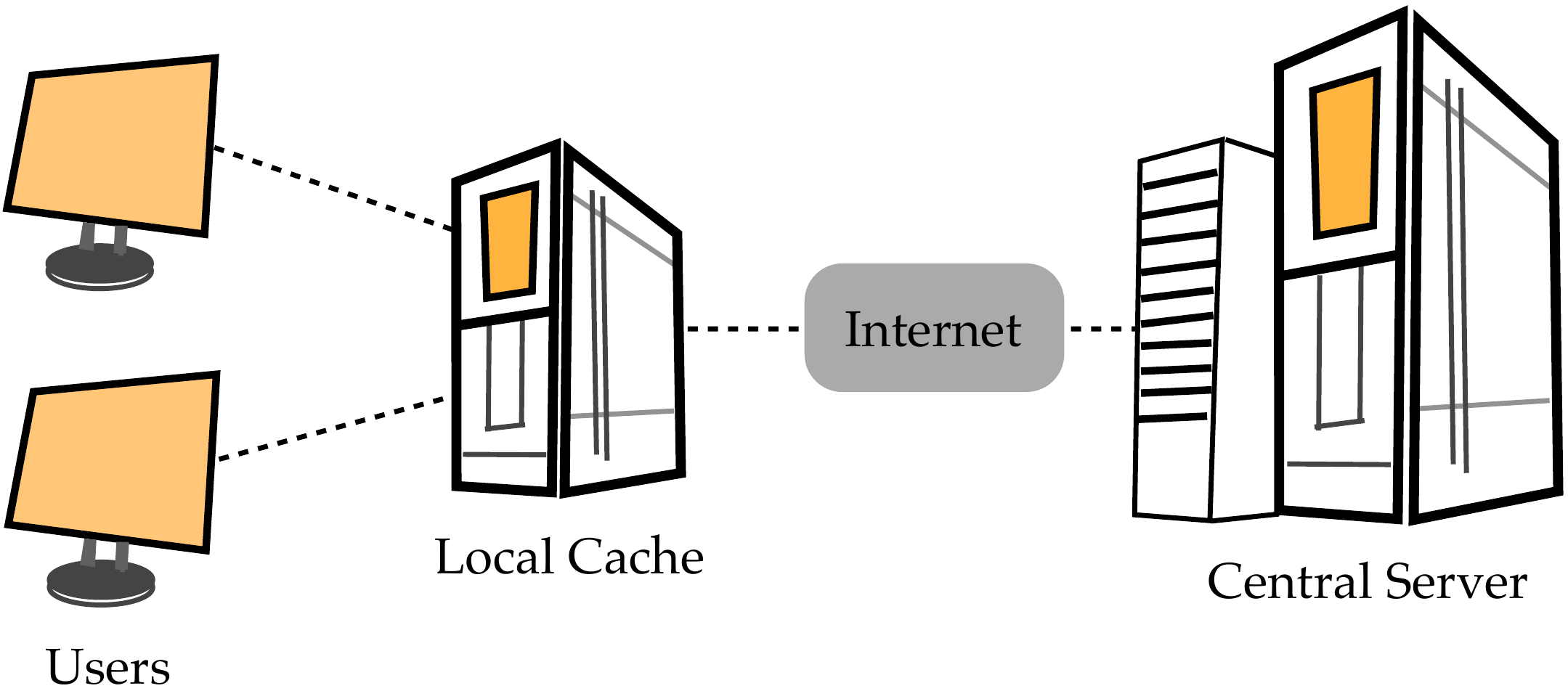}

\caption{An illustration of a Content Delivery Network (CDN) with a central server, a local cache and two users.}\label{fig:CDN_figure} 
\end{figure}

We consider a Content Delivery Network (CDN) consisting of a central server which stores the entire catalog of contents offered by the VoD service, assisted by a local cache with limited storage capacity (Figure \ref{fig:CDN_figure}). Content can be fetched from the central server and replicated on the local cache to serve user requests. The motivation behind such a network architecture is to serve most of the user requests via the local cache, thus reducing the load on the network backbone, and therefore reducing the overall bandwidth consumption of the network. In addition, the local cache can serve user requests with a lower start-up delay due to their geographical/network proximity. 

\subsection{Request Model}
We assume that the local cache serves $u$ users concurrently and the arrival requests from each user are generated i.i.d. according to the Markovian process described in Section \ref{section:our_model}. We assume that the service time of each request is an Exponential random variable with mean 1. 

\subsection{Cost Model}
\noindent We divide the cost of serving requests into two parts:
 \begin{enumerate}
 	\item[(i)] Cost of Bandwidth Usage: Each time a video is fetched from the central server and replicated on the local cache, the CDN pays a fetching cost denoted by $C_{\text{Fetch}}$.
 	\item[(ii)] Cost of Delayed Startup: Each time the requested content is not available in the local cache and has to be fetched from the central server \emph{after} the request is made, the CDN pays an additional start-up delay cost  denoted by $C_{\text{Delay}}$. This captures the cost of deterioration in the quality of service provided to the users.  
 \end{enumerate}
Without loss of generality, we normalize $C_{\text{Fetch}} = 1$ and let $C_{\text{Delay}} = \gamma \times C_{\text{Fetch}}$, where $\gamma$ is the start-up delay penalty.

Let $\text{Cost}(t)$ denote the total cost of serving requests that arrive before time $t$, $F(t)$ be the number of fetches from the central server to the local cache made before time $t$ and $D(t)$ be the number of delayed start-ups by time $t$. Then we have that,
$$\text{Cost}(t) = F(t) + \gamma \times D(t).$$

The goal is to design content caching policies to minimize the total cost of serving user requests. 

\section{Caching Policies}
\label{section:CachingPolicies}

We propose a caching policy which uses the fact that user requests are being generated according to a Markov process to determine which contents to cache. We refer to this policy as the PreFetch policy. The key idea of the PreFetch policy is to pre-fetch the top $r$ recommended videos as soon as a user requests a specific video, thus reducing the chance that the next request from this user will have to face any start-up delay. The policy uses the Least Recently Used (LRU) metric to purge stored content in order to make space to store the fetched content.

We use the following definitions in the formal definition of the PreFetch policy:
\begin{definition}
	\begin{enumerate}
		\item[--] A video is said to be \emph{in use} if it is being used to serve an active request.
		\item[--] A video is referred to as a \emph{tagged video} if it is one of the top $r$ (where $r$ is a pre-determined integer $\geq 1$) recommendations for any one the videos currently in use.
	\end{enumerate}
\end{definition}

Refer to Figure \ref{fig:alg_prefetch} for a formal definition of the PreFetch policy.

\begin{figure}[!hb]
	\hrule
	\vspace{0.1in}
	\begin{algorithmic}[1]
		\STATE \textbf{Input:} An integer $r \geq 1$.
		\STATE \textbf{Initialize:} Set of cached videos, $C = \phi$, set of tagged videos, $T = \phi$, set of videos in use, $U = \phi$, set of cached videos currently not in use or tagged, $V = C \setminus (T \cup U)$.
		\STATE On arrival (request for Video $i$) \textbf{do},
		\IF {Video $i \notin C$,}
		\IF {$|C| < \text{cache size}$,}
		\STATE fetch Video $i$; $C = C \cup \text {Video }i$
		\ELSIF {$V \neq \phi$,}
		\STATE fetch Video $i$; replace the Least Recently Used (LRU) video in $V$ with Video $i$.
		\ELSE
		\STATE remove a video $\in T$, chosen uniformly at random, and replace it with Video $i$.
		\ENDIF
		\STATE Update $C$, $V$, $T$ and $U$.
		\ENDIF
		\IF {top $r$ recommendations of Video $i$ not in cache,}
		\STATE pre-fetch missing recommended videos,
		\FOR {each pre-fetched video}
		\IF {$|C| < \text{cache size}$,}
		\STATE add video to the cache, update $C$,
		\ELSIF{$V \neq \phi$,}
		\STATE replace LRU video in $V$ with fetched video,
		\ELSE
		\STATE remove a video $\in T$, chosen uniformly at random, and replace it with fetched video.
		\ENDIF		
		\STATE Update $C$, $V$, $T$ and $U$.	
		\ENDFOR		
		\ENDIF
	\end{algorithmic}
	\vspace{0.1in}
	\hrule
	\caption{PreFetch -- \sl A caching policy which adapts the content stored on cache to ensure that the top $r$ recommended videos for the videos currently being viewed are pre-fetched to the cache in order to reduce the chance of start-up delay for the next request.} 
	\label{fig:alg_prefetch}
\end{figure}

\begin{remark}
	We assume that the storage capacity of the local cache is large enough to store more videos than the number of users it serves simultaneously. 
\end{remark}

\begin{remark}
	The PreFetch caching policy can be implemented without the knowledge of the relative popularity of various videos. The only information required to implement the PreFetch policy is the list of recommended videos corresponding to each video in the catalog, which is always known to the VoD service. 
\end{remark}

As discussed in \cite{khemmarat2012watching}, a possible generalization of the PreFetch policy is to pre-fetch only a fraction of the recommended videos instead of pre-fetching entire videos, and fetching the remaining part of the video only after the request is made. If there exists an $\alpha < 1$ such that while the user watches the first $\alpha$ fraction of the video, the remaining $(1-\alpha)$ fraction of the video can be pre-fetched, the CDN can provide uninterrupted service to the user without any start-up delay by pre-fetching only the first $\alpha$ fraction of the video.

In the next section, we compare the performance of our PreFetch policy with the popular Least Recently Used (LRU) caching policy. The LRU policy has been traditionally used for caching \cite{wang1999survey} and has been widely studied for decades. Refer to Figure \ref{fig:LRU} for a formal definition of the LRU policy.

\begin{figure}[!hb]
	\hrule
	\vspace{0.1in}
	\begin{algorithmic}[1]
		\STATE On arrival (request for Video $i$) \textbf{do},
		\IF {Video $i$ not present in the cache,}
		\STATE fetch Video $i$; replace the Least Recently Used (LRU) cached video with Video $i$.
		\ENDIF
	\end{algorithmic}
	\vspace{0.1in}
	\hrule
	\caption{Least Recently Used (LRU) -- \sl A caching policy.} 
	\label{fig:LRU}
\end{figure}
\section{Simulation Results}
\label{section:simulate}
In this section, we compare the performance of the LRU policy and the PreFetch policy. Our goal is to understand if exploiting the time correlation between requests from a user by pre-fetching recommended videos can lead to better performance. In addition, we also study how the performance of the two caching policies depends on the request arrival process and various system parameters like number of users using a local server ($u$), size of cache, fraction of video pre-fetched ($\alpha$).

Requests arrive according to the request model discussed in Section \ref{section:our_model}. We assume that the VoD service has a content catalog consisting of 1000 videos. We use the Albert-Barabasi model (Figure \ref{fig:BA}) to generate the recommendation graph with $m = 20$. We fix $\kappa = 0.8$ (defined in Section \ref{section:our_model}) for all the results presented in this section. We assume that the service time of each request is an Exponential random variable with mean of one time unit. We assume all videos are of unit size. For each set of system parameters, we simulate the system for $10^5$ time units. 

\subsection{Cost v/s Startup delay penalty  $(\gamma)$ }

\begin{figure}[!hb]
\centering
\includegraphics[width = 0.7\textwidth]{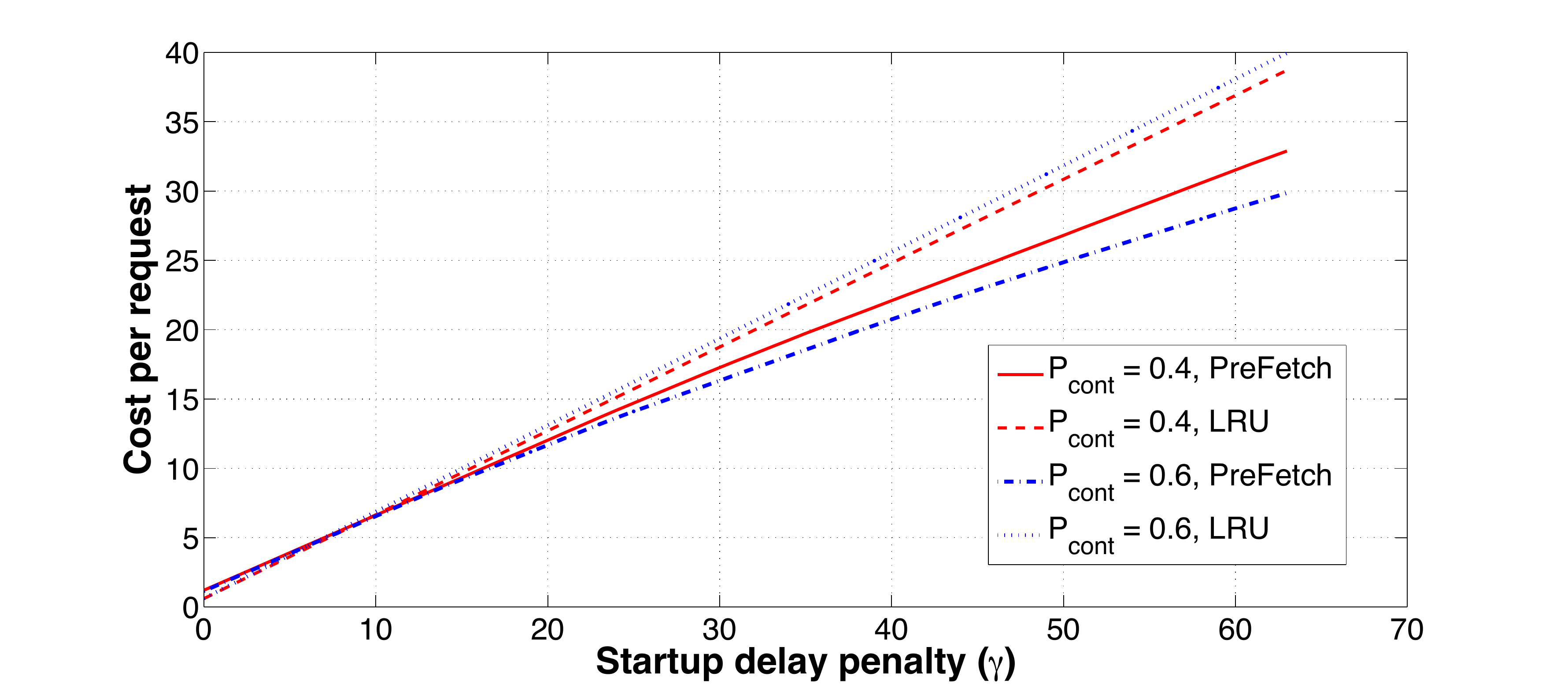}

\caption{Cost vs Start-up delay penalty ($\gamma$) for a system with Number of videos $= 1000$, $m = 20$, Zipf parameter $(\beta)  = 0.8$, Cache size $= 200$ and $1$ User. As $\gamma$ increases, PreFetch outperforms the LRU policy. }\label{fig:CostvsW} 
\end{figure}

\begin{figure}[!hb]
\centering
\includegraphics[width = 0.7\textwidth]{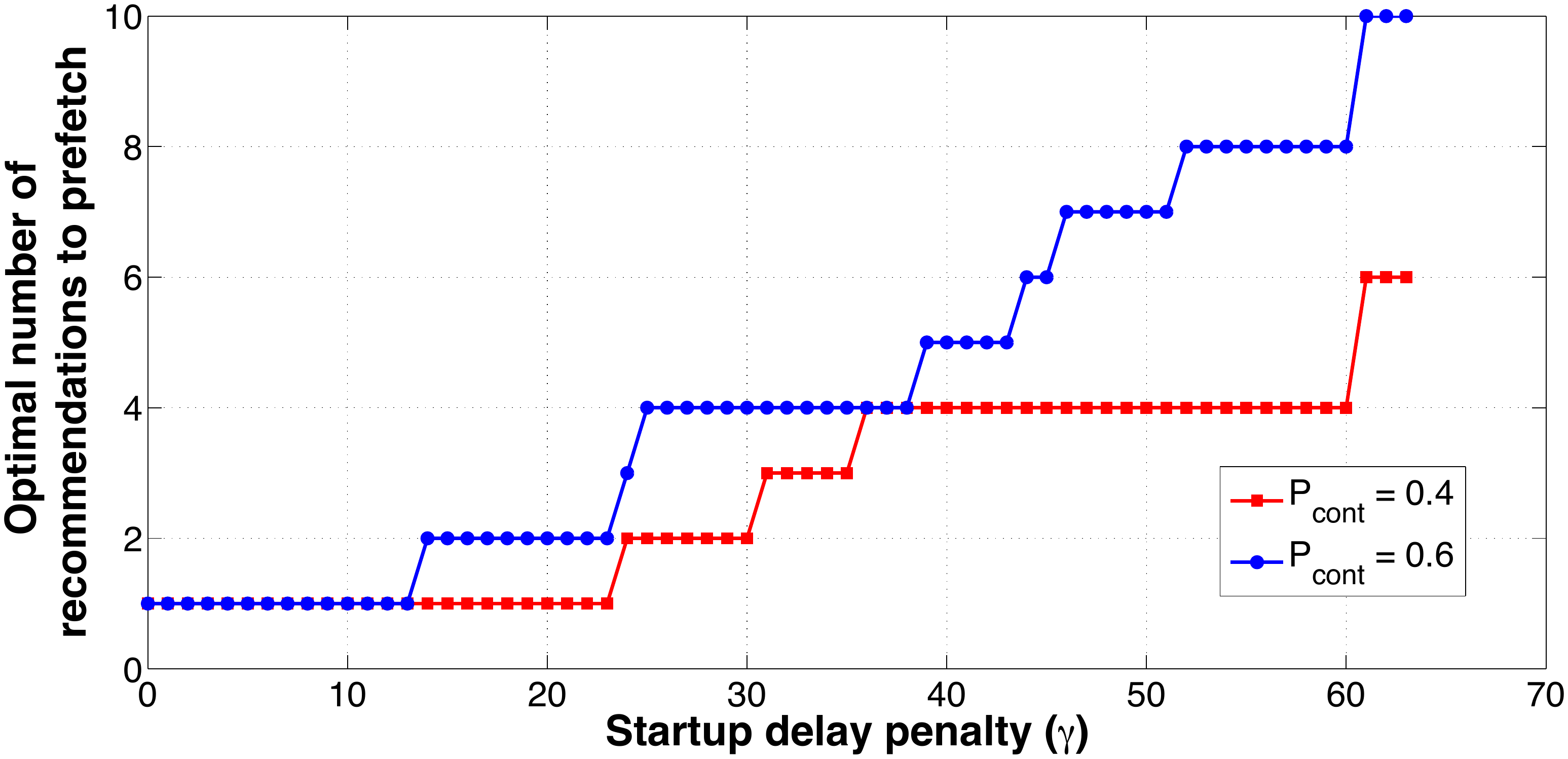}

\caption{The optimal number of recommendations to pre-fetch $(r)$ vs Start-up delay penalty ($\gamma$) for a system with Number of videos $= 1000$, $m = 20$, Zipf parameter $(\beta)  = 0.8$, Cache size $= 200$ and $1$ User. The optimal number of recommendations to pre-fetch ($r$) increases with start-up delay penalty $\gamma$. }\label{fig:RecosvsW} 
\end{figure}

In Figure \ref{fig:CostvsW}, we compare the performance of the PreFetch policy and the LRU policy as a function of the Start-up delay penalty ($\gamma$). Recall that $P_{cont}$ is the probability that the next video requested by the user is one of the recommended videos. The PreFetch policy pre-fetches the top $r$ $(\geq 1)$ recommendations of a video from the central server to the cache the moment a video is requested, thus ensuring that there is no start-up delay if the user requests one the top $r$ recommended videos. In addition, the total cost of service is the sum of the cost of bandwidth usage and cost due to startup delay. In Figure \ref{fig:CostvsW}, for each value of Start-up delay penalty ($\gamma$), we use the empirically optimized value of $r$ which leads to the lowest cost of service. The optimal value of the number of recommendations to pre-fetch $(r)$ increases with increase in Start-up delay penalty $(\gamma)$ as shown in Figure \ref{fig:RecosvsW}. 

We observe that for low values of Start-up delay penalty $(\gamma)$, LRU outperforms the PreFetch policy. As the Start-up delay penalty $(\gamma)$ increases, PreFetch outperforms the LRU policy. This illustrates the tradeoff between bandwidth usage, i.e., number of pre-fetches and quality of service, i.e., reducing startup delay. 
 

\subsection{Cost v/s Number of users $(u)$}
\begin{figure}[!hb]
\centering
\includegraphics[width = 0.7\textwidth]{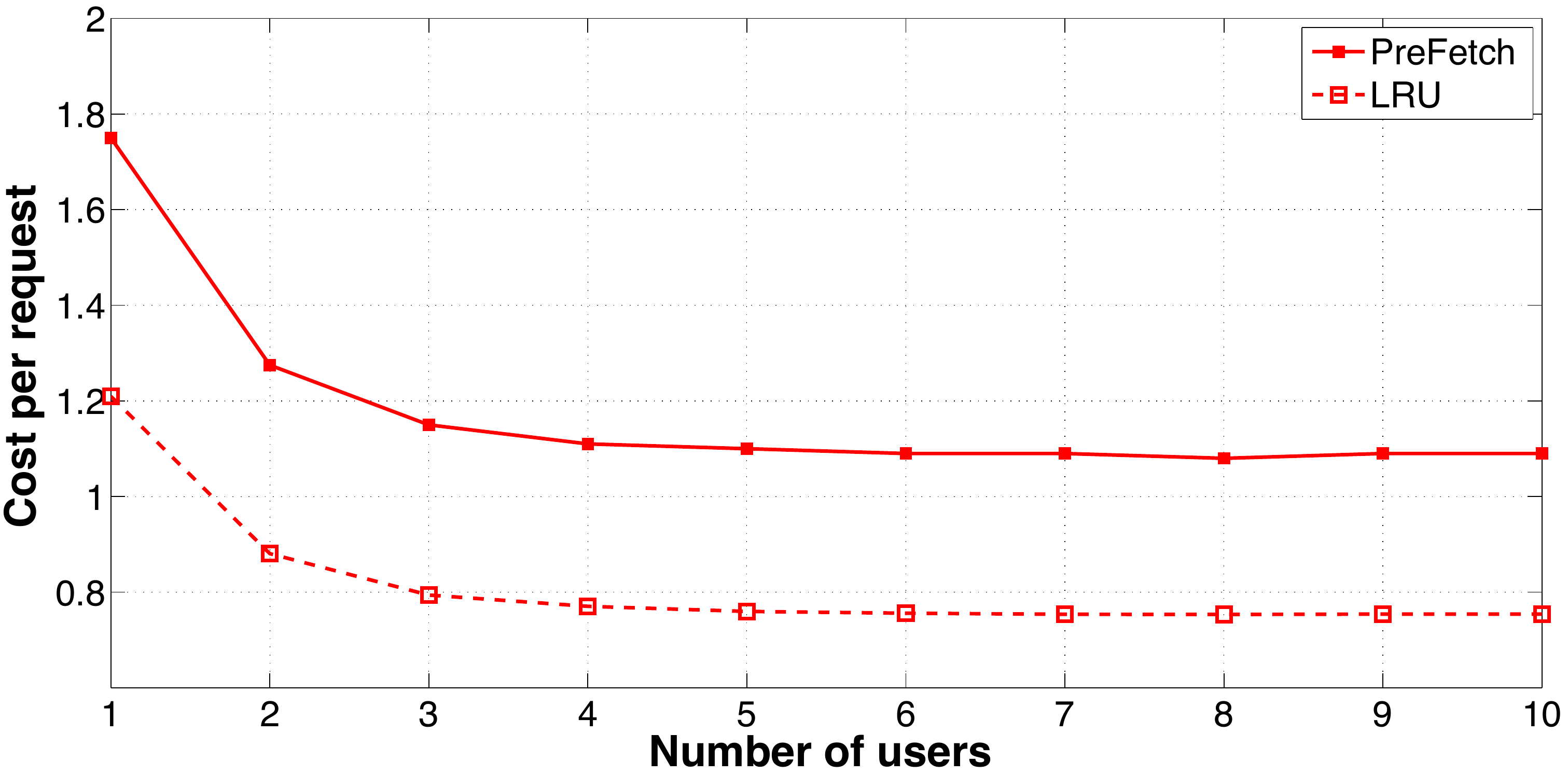}

\caption{Cost vs Number of users for a system with Number of videos $= 1000$, Startup delay penalty $(\gamma) = 1$, Zipf parameter $(\beta)  = 0.8$, $P_{cont} = 0.4$ and Cache size $= 200$. LRU outperforms the PreFetch policy. }\label{fig:CostvsUsers2} 
\end{figure}

\begin{figure}[h]
\centering
\includegraphics[width = 0.7\textwidth]{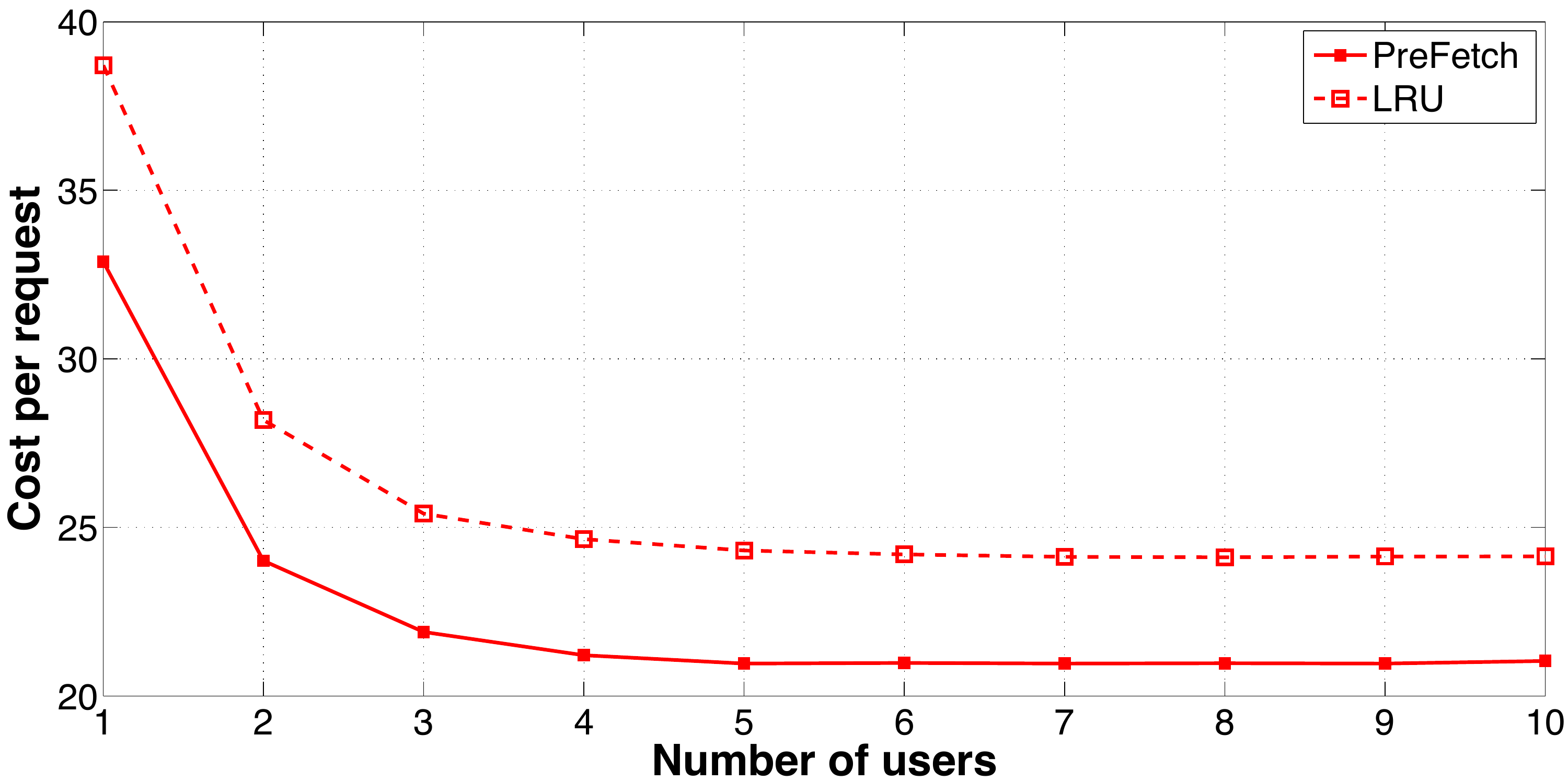}

\caption{Cost vs Number of users for a system with Number of videos $= 1000$, Starup delay penalty $(\gamma) = 63$, Zipf parameter $(\beta) = 0.8$, $P_{cont} = 0.4$ and Cache size $= 200$. The PreFetch policy outperforms the LRU policy. }\label{fig:CostvsUsers64} 
\end{figure}

\begin{figure}[h]
\centering
\includegraphics[width = 0.7\textwidth]{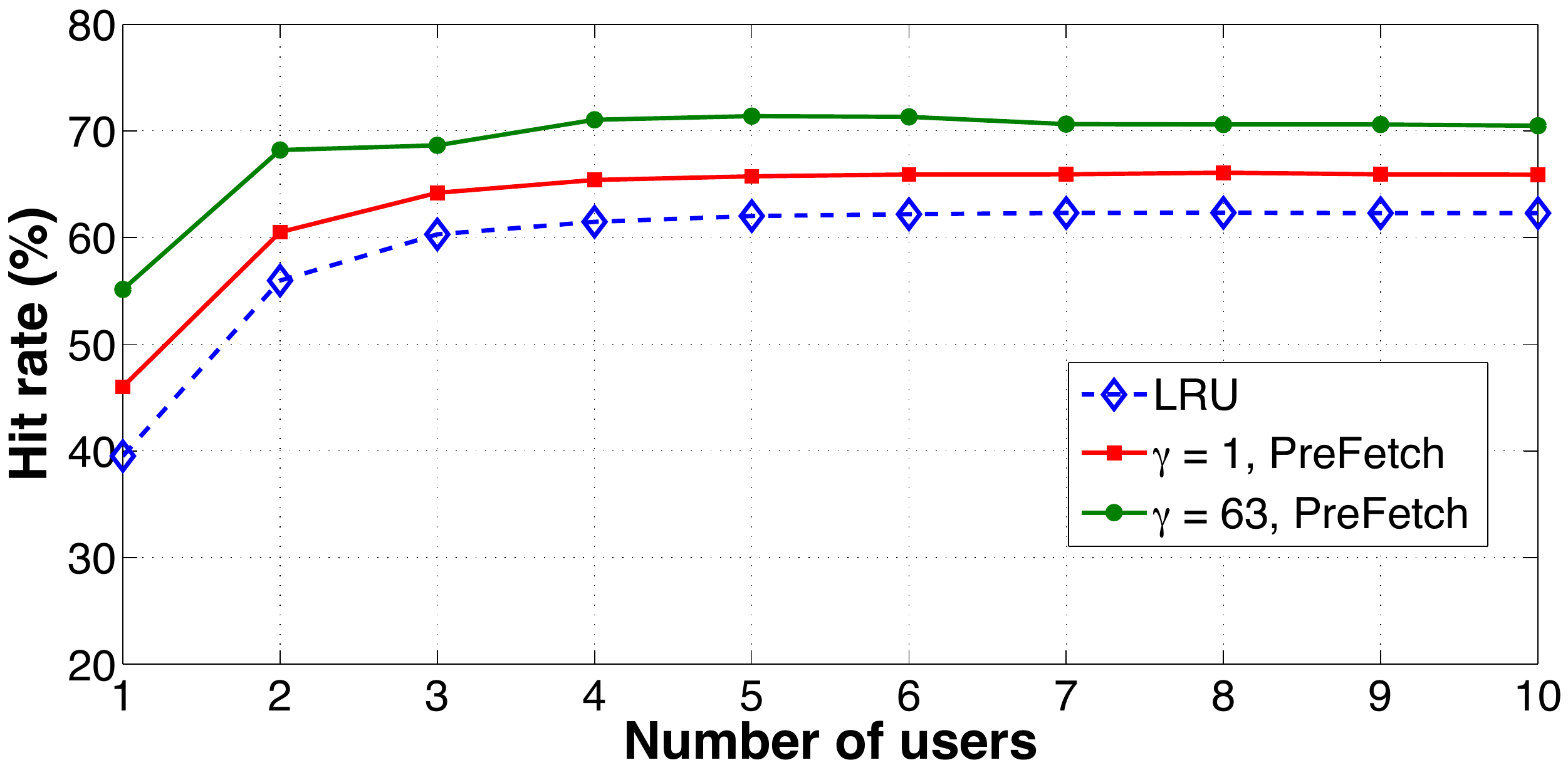}

\caption{Hit rate vs Number of users for a system with Number of videos $= 1000$, Zipf parameter $(\beta)  = 0.8$, $P_{cont} = 0.4$ and Cache size $= 200$. Cache hit rates are significantly improved by using PreFetch scheme for all values of $u$ and $\gamma$. }\label{fig:HitRatevsNumOfUsers} 
\end{figure}


In Figures \ref{fig:CostvsUsers2} and \ref{fig:CostvsUsers64}, we compare the performance of the two policies where the value of $r$ used by the PreFetch policy is empirically optimized for each value of $u$ and $\gamma$. We see that as the number of users increases from 1 to 5, there is a sharp drop in the cost for both LRU and PreFetch policy. Since all the users access videos according to the same Markov process, when there are multiple users accessing the cache, the probability that the popular videos and their top recommendations are always present in the cache increases. This reduces the number of cache misses and the number of pre-fetches for the most popular videos, thus reducing the overall cost of service. 

As seen in Figure \ref{fig:CostvsUsers2}, when the Startup delay cost $\gamma$ is low, the LRU caching policy outperforms the (optimized) PreFetch policy for all values of $u$. For $\gamma = 63$ (Figure \ref{fig:CostvsUsers64}), the PreFetch policy outperforms the LRU caching policy. 
Our simulations shows that for $\gamma \leq 11$, the optimal number of recommendations ($r$) to cache is 1. The optimal value of $r$ is between 4 - 6 for $\gamma = 63$. 

Figure \ref{fig:HitRatevsNumOfUsers} illustrates that cache hit rates are higher for the PreFetch policy as compared to that of the LRU policy for all values of $u$ and $\gamma$ considered.

\subsection{Cost v/s $P_{cont}$}

\begin{figure}[h]
\centering
\includegraphics[width = 0.7\textwidth]{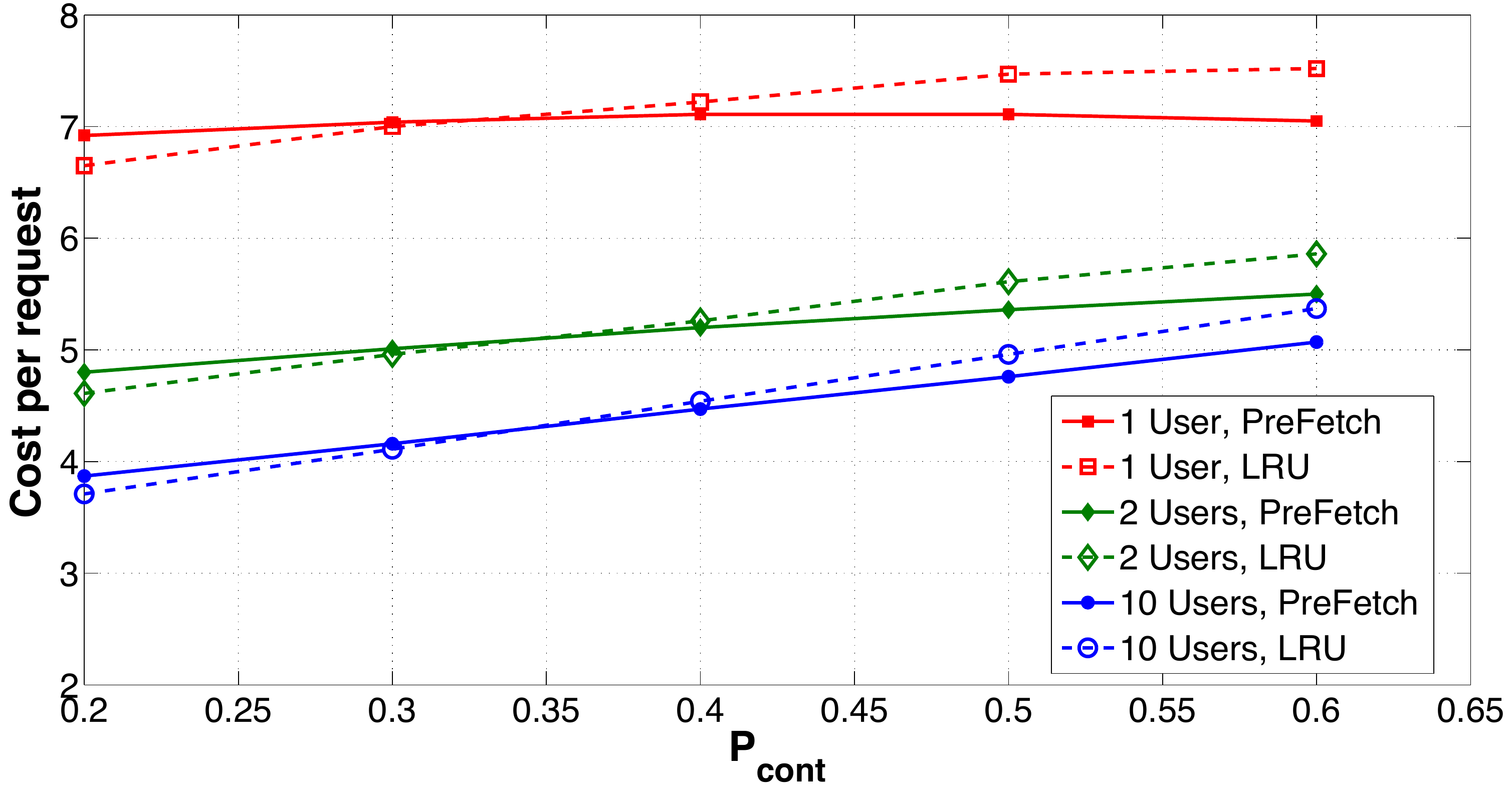}

\caption{Cost vs $P_{cont}$ for a system with Number of videos $= 1000$, Startup delay penalty $(\gamma) = 11$, Zipf parameter $(\beta)  = 0.8$ and Cache size $= 200$. For low values of $P_{cont}$, the excess bandwidth usage due to pre-fetching outweighs the benefits of reducing startup delay, and for higher values of $P_{cont}$, pre-fetching leads to reduced cost of service. }\label{fig:CostvsPcont12} 
\end{figure}

\begin{figure}[h]
\centering
\includegraphics[width = 0.7\textwidth]{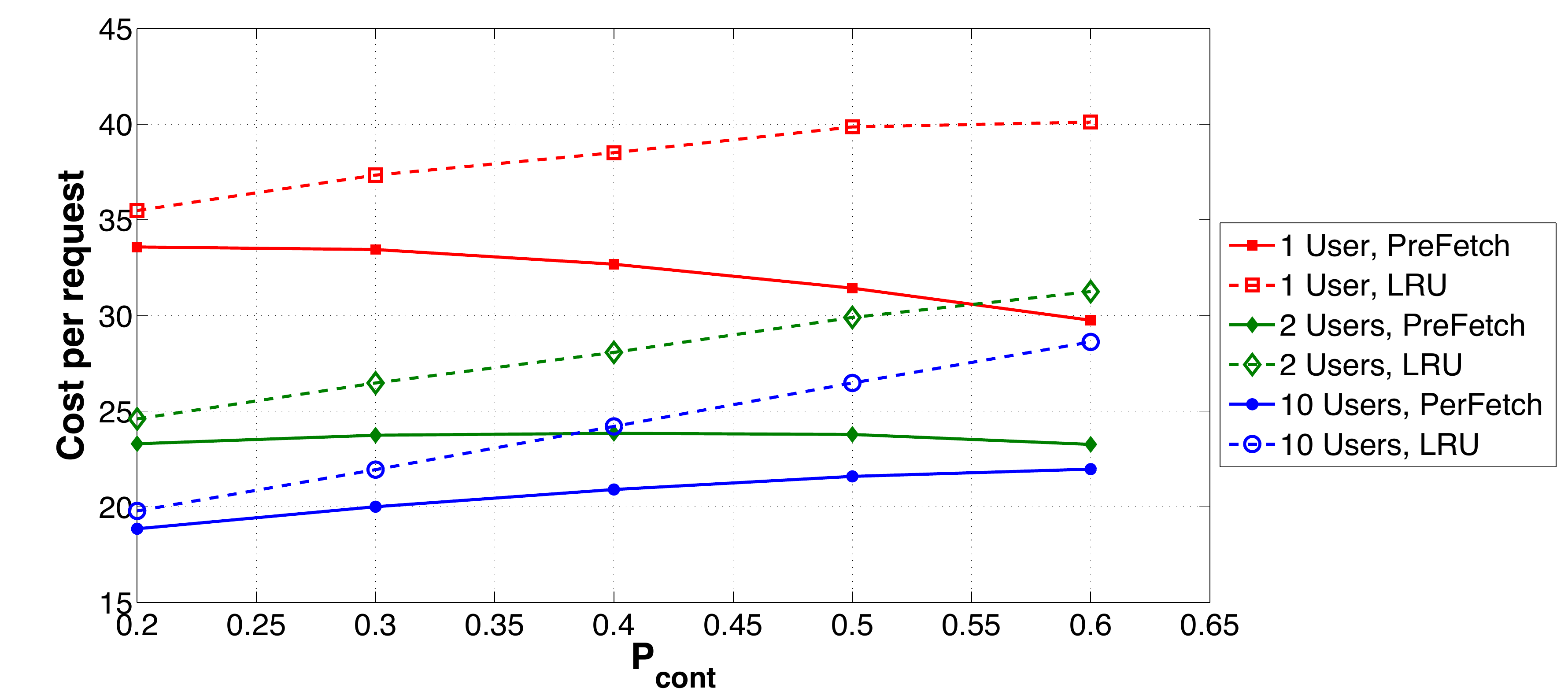}

\caption{Cost vs $P_{cont}$ for a system with Number of videos $= 1000$, Startup delay cost $(\gamma) = 63$, Zipf parameter $(\beta) = 0.8$ and Cache size $= 200$. The PreFetch policy outperforms LRU for all values of $P_{cont}$ considered. }\label{fig:CostvsPcont64} 
\end{figure}
 
Recall that $P_{cont}$ denotes the probability that the next video is accessed via the recommendation list. We vary the value of $P_{cont}$ between $0.2$ and $0.6$ (to be consistent with the observations in \cite{krishnappa2015cache}) and evaluate the performance of LRU and optimal PreFetch policy for $\gamma = 11$ and $\gamma = 63$. 

In Figure \ref{fig:CostvsPcont12}, we see that LRU outperforms the (optimized) PreFetch policy for low values of $P_{cont}$ and PreFetch outperforms LRU as $P_{cont}$ increases. Since increasing the value of $P_{cont}$ increases the probability that the next video is accessed via the recommendation list, we conclude that if the Startup delay cost is not very high ($\gamma = 11$), for low values of $P_{cont}$, the excess bandwidth usage due to pre-fetching outweighs the benefits of reducing startup delay. 

Figure \ref{fig:CostvsPcont64} illustrates that the PreFetch policy outperforms LRU for $\gamma = 63$ for all values of $P_{cont}$ considered. In addition, the relative performance of PreFetch policy improves with respect to LRU policy with increase in $P_{cont}$. 

\begin{figure}[h]
\centering
\includegraphics[width = 0.7\textwidth]{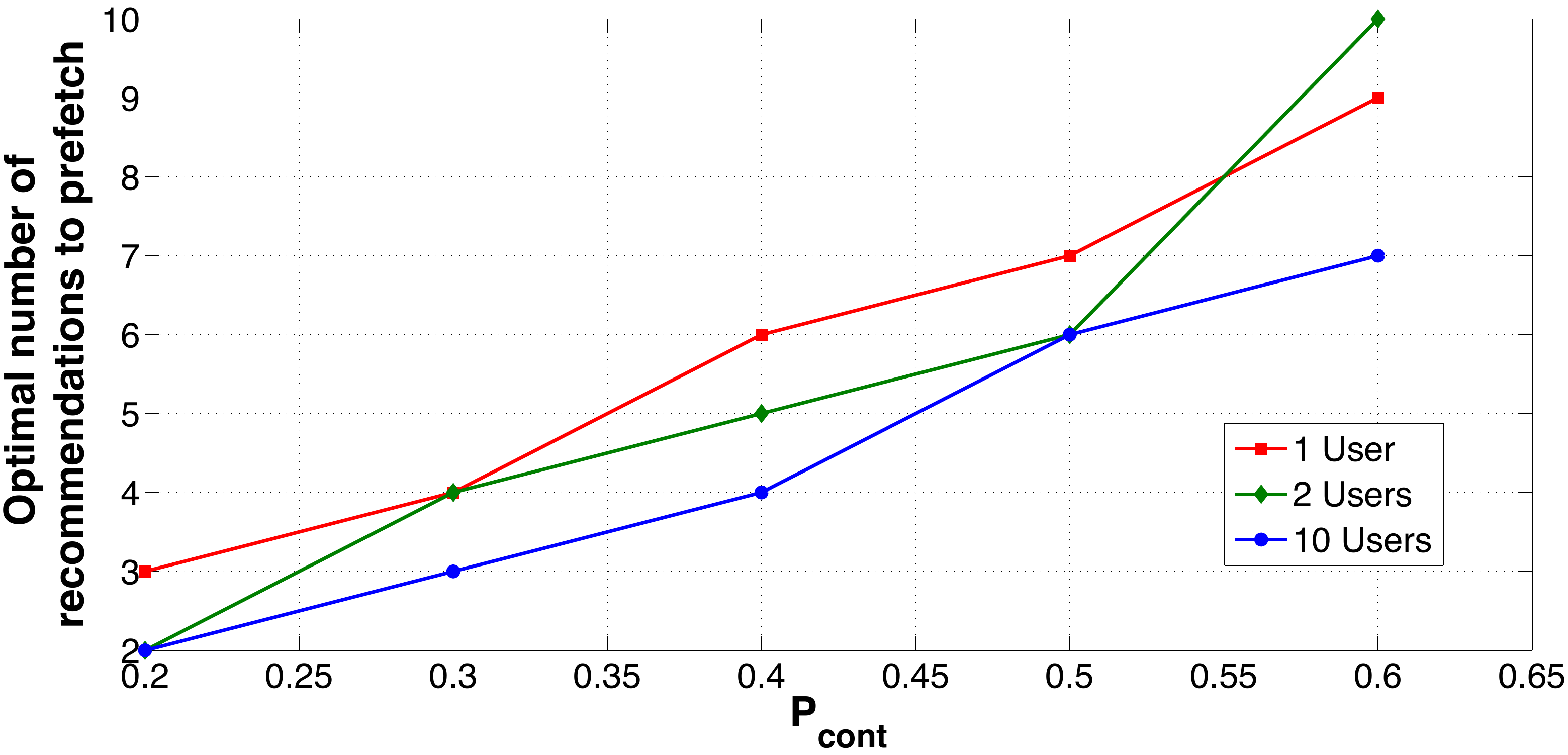}

\caption{Optimal number of recommendations to pre-fetch $(r)$ vs $P_{cont}$ for a system with Number of videos $= 1000$, Startup delay penalty $(\gamma) = 63$, Zipf parameter $(\beta)  = 0.8$ and Cache size $= 200$. The optimal number of recommendations to pre-fetch increases with $P_{cont}$. }\label{fig:CostvsRecos64} 
\end{figure}

\begin{figure}[h]
\centering
\includegraphics[width = 0.7\textwidth]{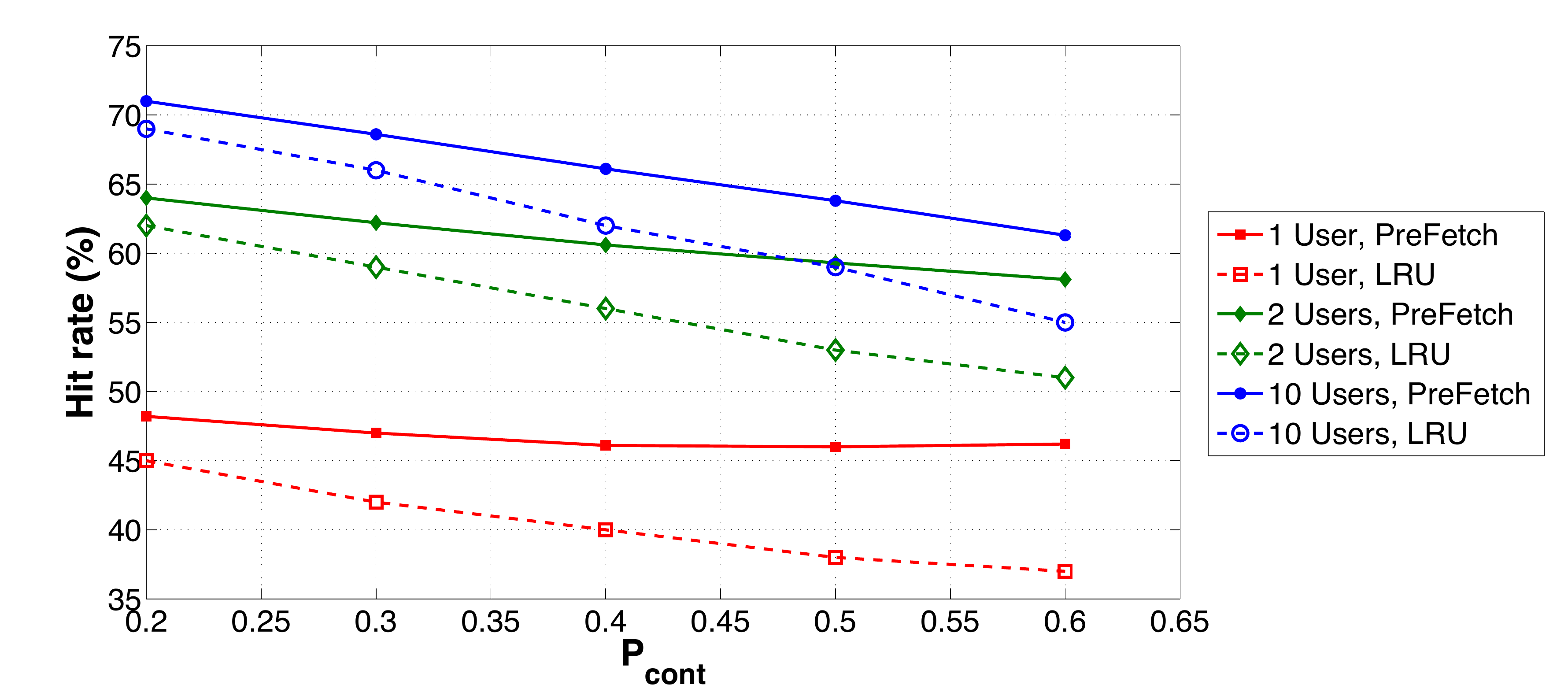}

\caption{Hit rate vs $P_{cont}$ for a system with Number of videos $= 1000$, Startup delay penalty $(\gamma) = 11$, Zipf parameter $(\beta) = 0.8$ and Cache size $= 200$. The PreFetch policy has higher hit rate and the difference between the hit rates of the PreFetch policy and the LRU policy increases with increasing $P_{cont}$. }\label{fig:HitvsPcont12} 
\end{figure}

\begin{figure}[h]
\centering
\includegraphics[width = 0.7\textwidth]{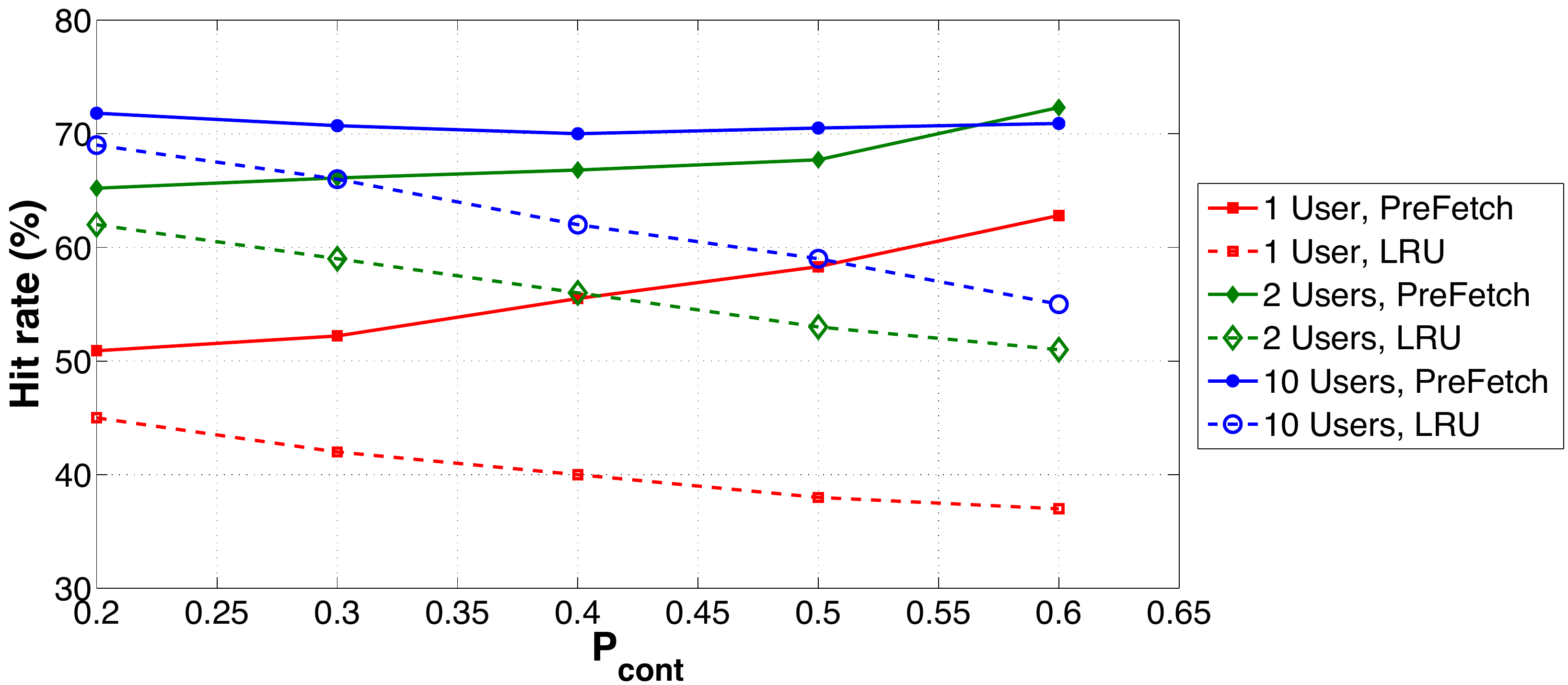}

\caption{Hit rate vs $P_{cont}$ for a system with Number of videos $= 1000$, Startup delay penalty $(\gamma) = 63$, Zipf parameter $(\beta) = 0.8$ and Cache size $= 200$. The PreFetch policy has higher hit rate and the difference between the hit rates of the PreFetch policy and the LRU policy increases with increasing $P_{cont}$.}\label{fig:HitvsPcont64} 
\end{figure}
In Figure \ref{fig:CostvsRecos64}, we plot the optimal value of $r$ as a function of $P_{cont}$. We conclude that with increasing $P_{cont}$, it is beneficial to pre-fetch more videos from the recommendation list.

Figures \ref{fig:HitvsPcont12} and \ref{fig:HitvsPcont64}, corresponding to $\gamma = 11$ and $\gamma = 63$ respectively, illustrate that cache hit rate is higher for the PreFetch policy as compared to the LRU policy. 

\subsection{Cost v/s Zipf parameter $(\beta)$}

\begin{figure}[H]
\centering
\includegraphics[width = 0.7\textwidth]{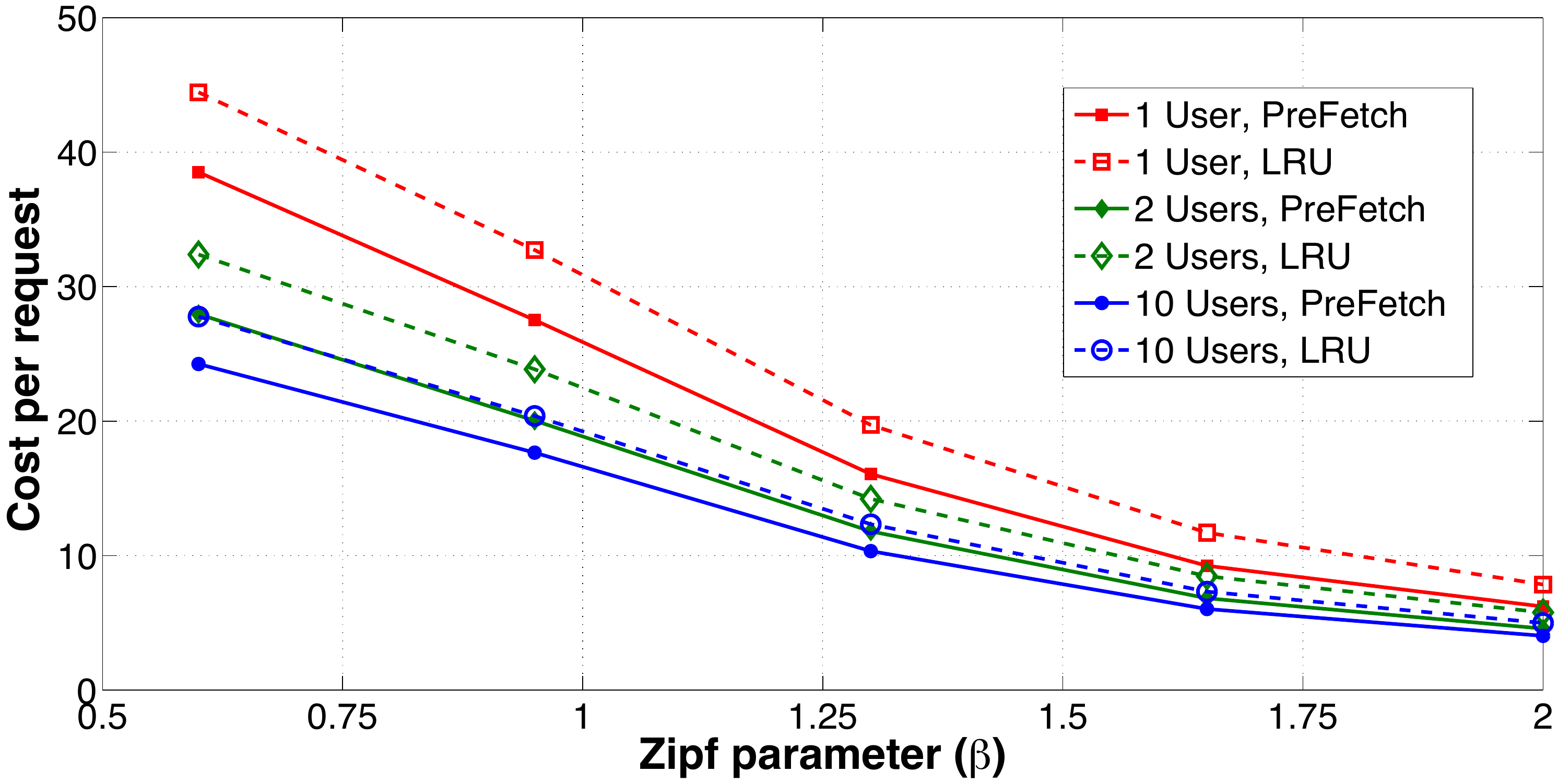}

\caption{Cost vs Zipf parameter $(\beta)$ for a system with Number of videos $= 1000$, Start-up delay penalty $(\gamma) = 63$, $P_{cont}  = 0.4$ and Cache size $= 200$. The performance of both the LRU policy and the PreFetch policy improve with increasing $\beta$. }\label{fig:CostvsBeta64} 
\end{figure}

\begin{figure}[h]
\centering
\includegraphics[width = 0.7\textwidth]{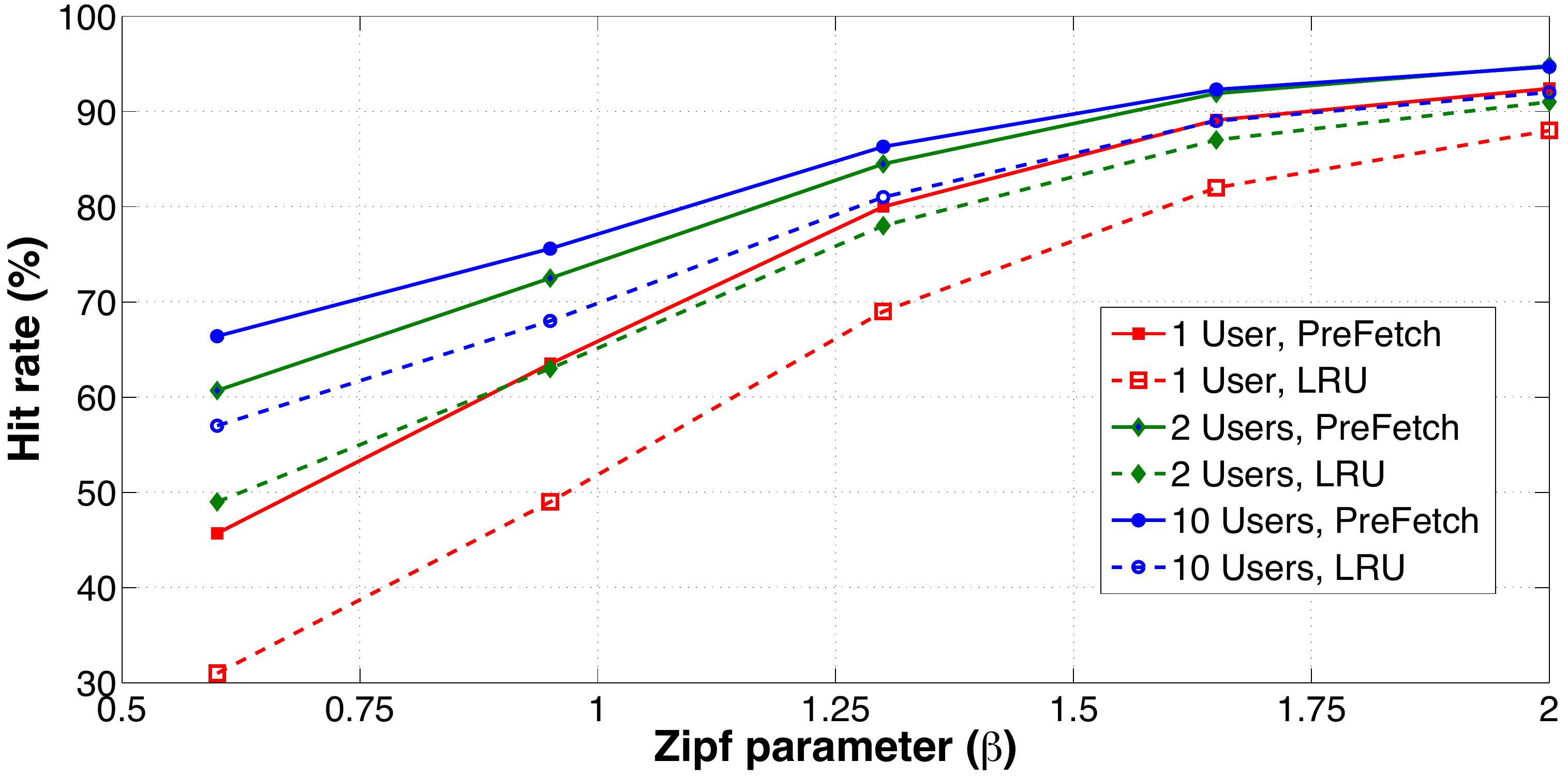}

\caption{Hit rate vs Zipf parameter $(\beta)$ for a system with Number of videos $= 1000$, Start-up delay penalty $(\gamma) = 63$, $P_{cont}  = 0.4$ and Cache size $= 200$. The hit rates for both the LRU policy and the PreFetch policy improve with increasing $\beta$. }\label{fig:HitvsBeta} 
\end{figure}

As discussed in Figure \ref{fig:final_popularity}, increasing the value of the Zipf parameter $\beta$ makes the overall content popularity more lopsided, i.e., a smaller fraction of the videos account for the same fraction of the total requests. Therefore, the performance for both the LRU policy and the PreFetch policy improves with increasing $\beta$ (Figure \ref{fig:CostvsBeta64}), as the small pool of popular videos are available in the local cache more often for both policies. We focus on $\beta$ values between 0.6 and 2 since typical values of $\beta$ lie in that range for most VoD services \cite{liu2013measurement,liu2012server,breslau1999web,yu2006understanding,iamnitchi2004small,
veloso2002hierarchical,fricker2012impact}. For Startup delay penalty $\gamma > 11$, the PreFetch policy outperforms the LRU policy for all $\beta$ between $0.6 - 2$.  Optimal $r$ for $\gamma = 63$ falls between $ 4 - 6$ for these values of $\beta$. Figure \ref{fig:HitvsBeta} illustrates that cache hit rates increase with increasing $\beta$.


\subsection{Cost v/s Cache size}
\begin{figure}[h]
\centering
\includegraphics[width = 0.7\textwidth]{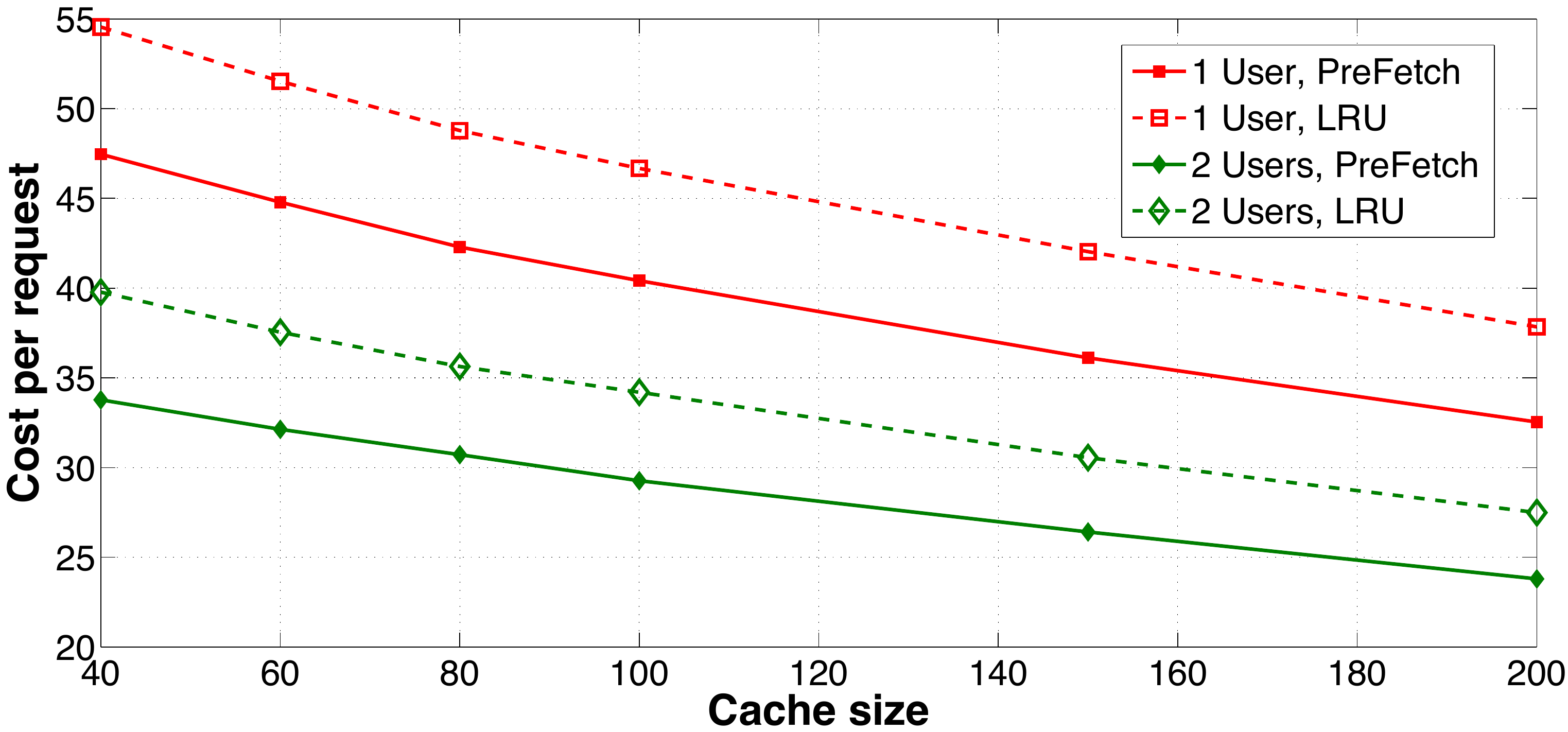}

\caption{Cost vs Cache size for a system with Number of videos $= 1000$, Startup delay penalty $(\gamma) = 63$, $P_{cont}  = 0.4$ and Zipf parameter $(\beta) = 0.8$. The performance of both polices improves with increasing cache size. }\label{fig:CostvsCache64} 
\end{figure}

\begin{figure}[h]
\centering
\includegraphics[width = 0.7\textwidth]{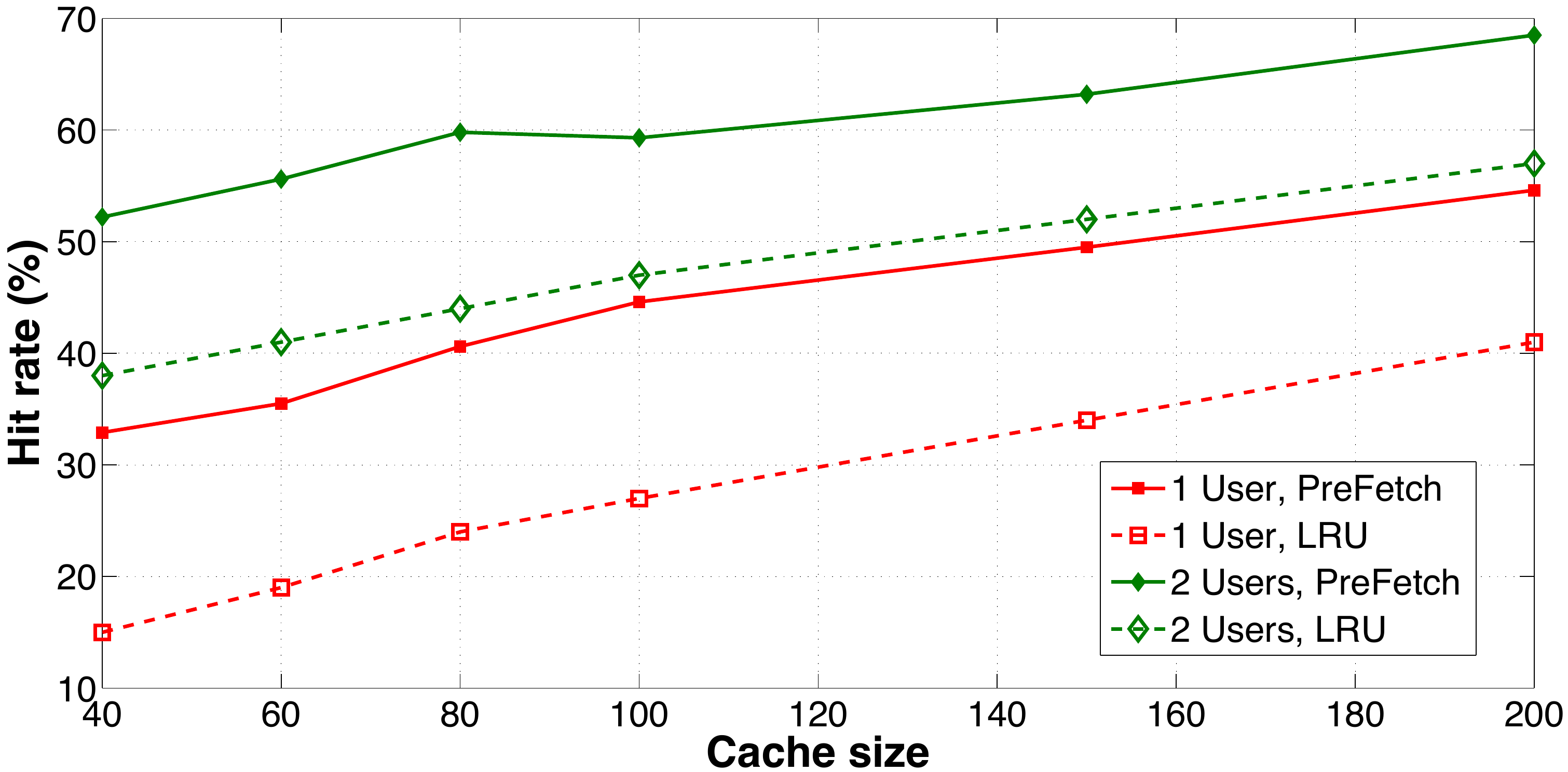}

\caption{Hit rate vs Cache size for a system with Number of videos = $1000$, Startup delay cost $(\gamma) = 63$, $P_{cont}  = 0.4$ and Zipf parameter $(\beta) = 0.8$. The hit rates for both polices improve with increasing cache size. }\label{fig:HitvsCache64} 
\end{figure}

We expect the performance of all policies to improve with the increase in cache size. In Figures \ref{fig:CostvsCache64} and \ref{fig:HitvsCache64}, we see that the PreFetch policy performs considerably better than the LRU policy for all cache sizes considered.

\subsection{Cost v/s Fraction to prefetch $(\alpha)$}

\begin{figure}[h]
\centering
\includegraphics[width = 0.7\textwidth]{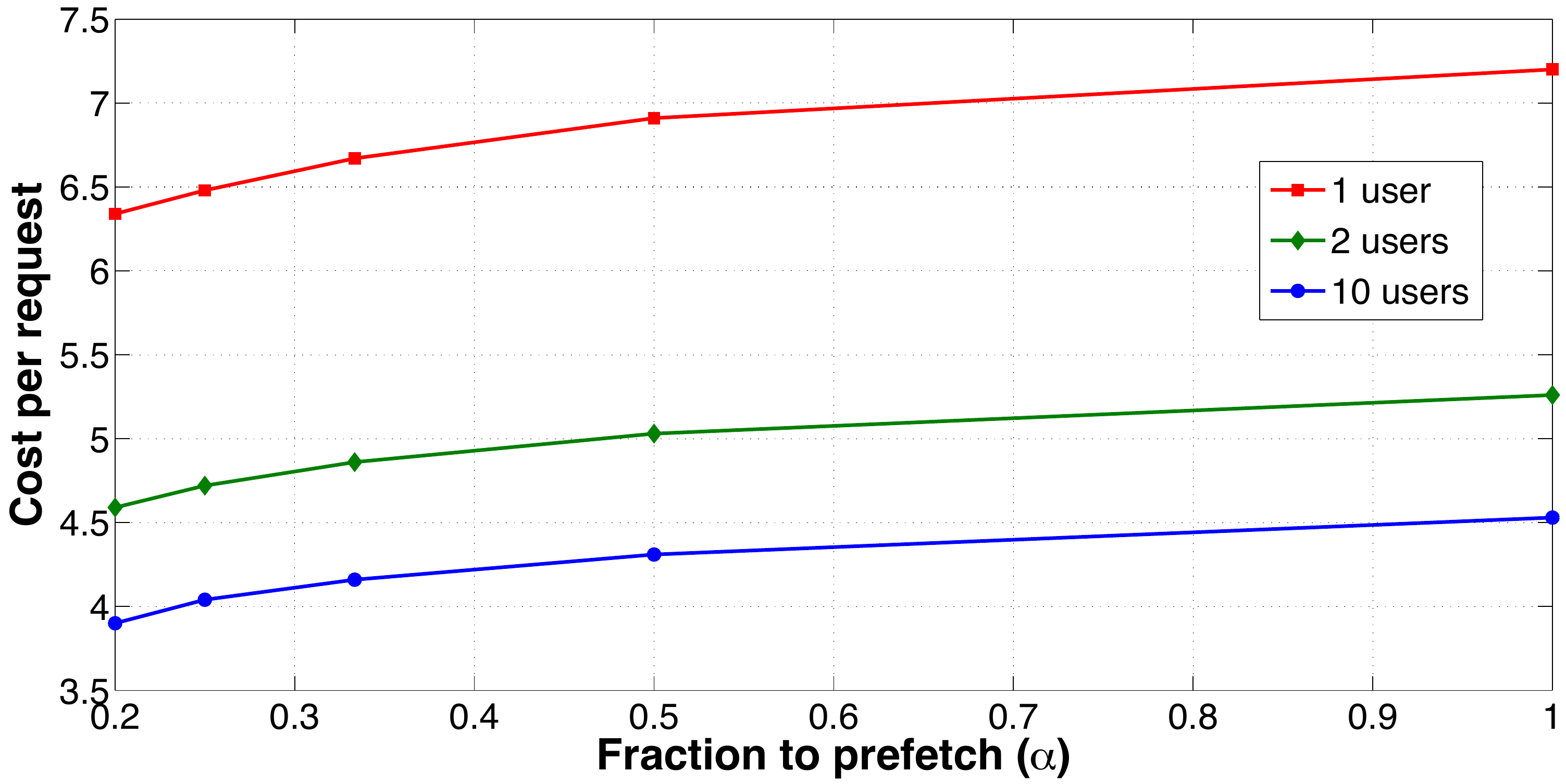}

\caption{Cost vs Fraction to pre-fetch $(\alpha)$ for a system with Number of videos $= 1000$, Startup delay penalty $(\gamma) = 11$, $P_{cont}  = 0.4$, Zipf parameter $(\beta) = 0.8$ and Cache size $= 200$. The cost of service increases with $\alpha$ as larger fractions of videos need to be pre-fetched to avoid start-up delay. }\label{fig:CostvsAlpha12} 
\end{figure}

\begin{figure}[h]
\centering
\includegraphics[width = 0.7\textwidth]{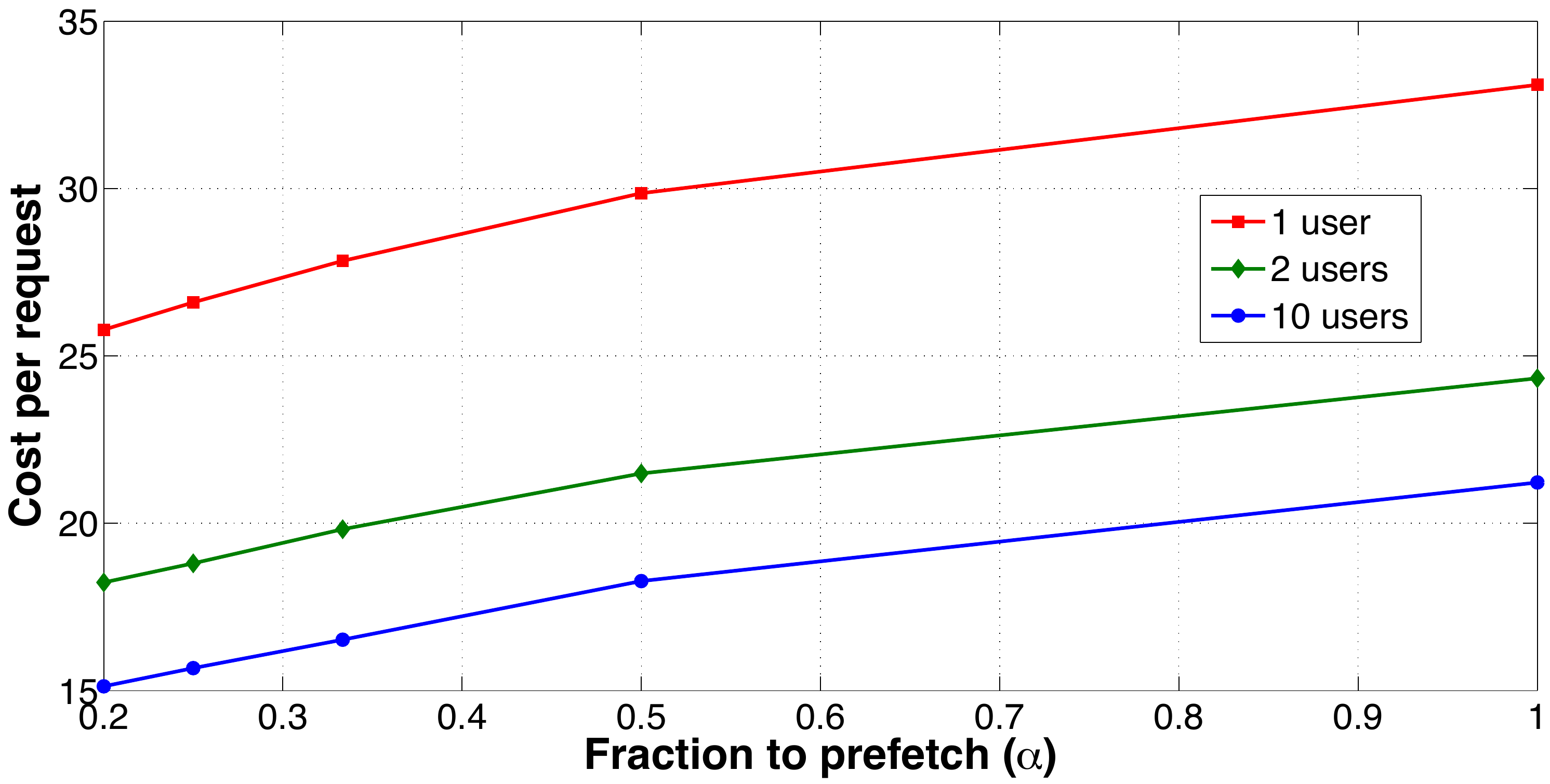}

\caption{Cost vs Fraction to pre-fetch $(\alpha)$ for a system with Number of videos $= 1000$, Startup delay penalty $(\gamma) = 63$, $P_{cont}  = 0.4$, Zipf parameter $(\beta) = 0.8$ and Cache size $= 200$. The cost of service increases with $\alpha$ as larger fractions of videos need to be pre-fetched to avoid start-up delay. }\label{fig:CostvsAlpha64} 
\end{figure}

\begin{figure}[h]
\centering
\includegraphics[width = 0.7\textwidth]{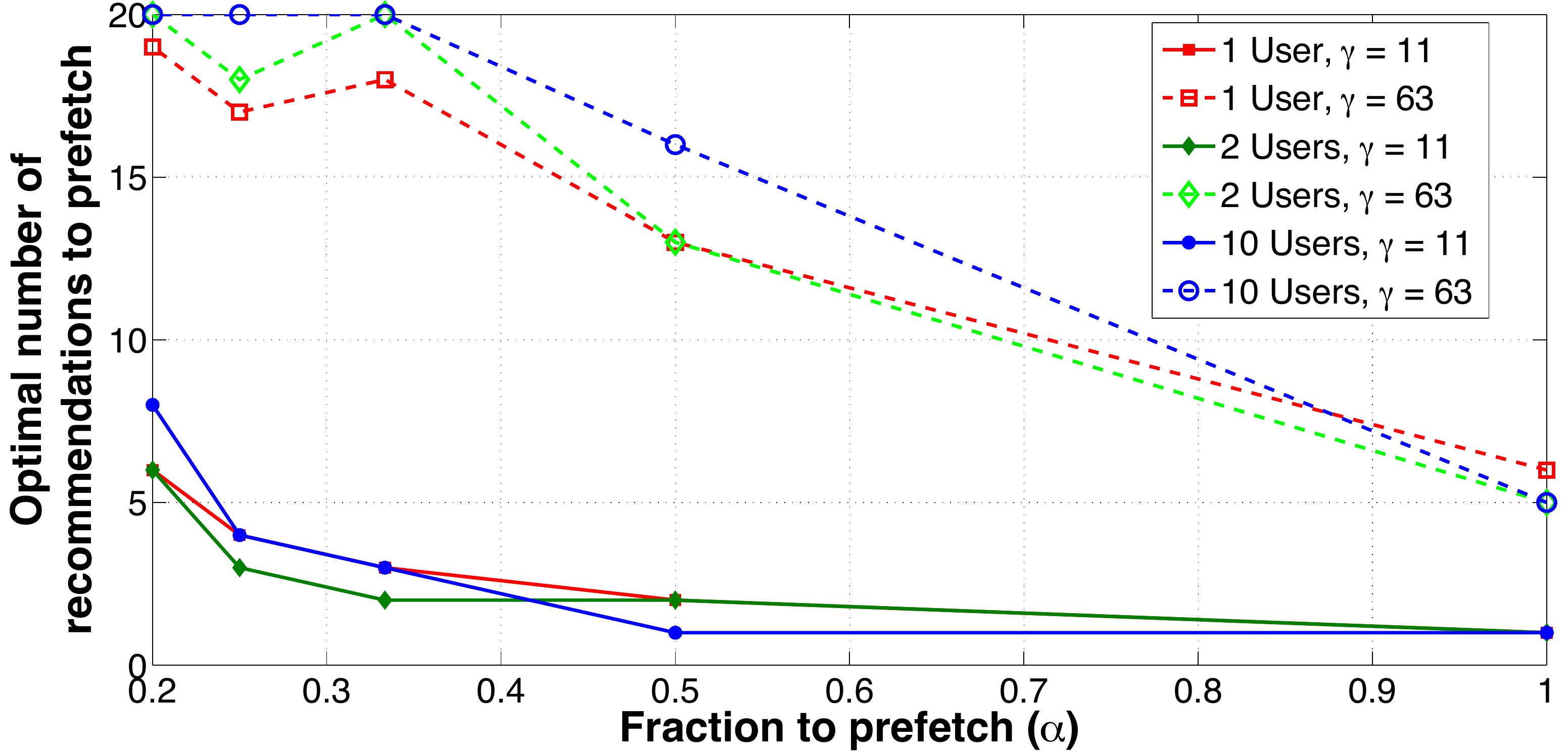}

\caption{Optimal number of recommendations to pre-fetch $(r)$ vs Fraction to pre-fetch $(\alpha)$ for a system with Number of videos $= 1000$, $P_{cont}  = 0.4$, Zipf parameter $(\beta) = 0.8$ and Cache size $= 200$. The optimal number of recommendations to pre-fetch decrease with $\alpha$ as larger fractions of videos need to be pre-fetched to avoid start-up delay.  }\label{fig:RecosvsAlpha} 
\end{figure}

\begin{figure}[h]
\centering
\includegraphics[width = 0.7\textwidth]{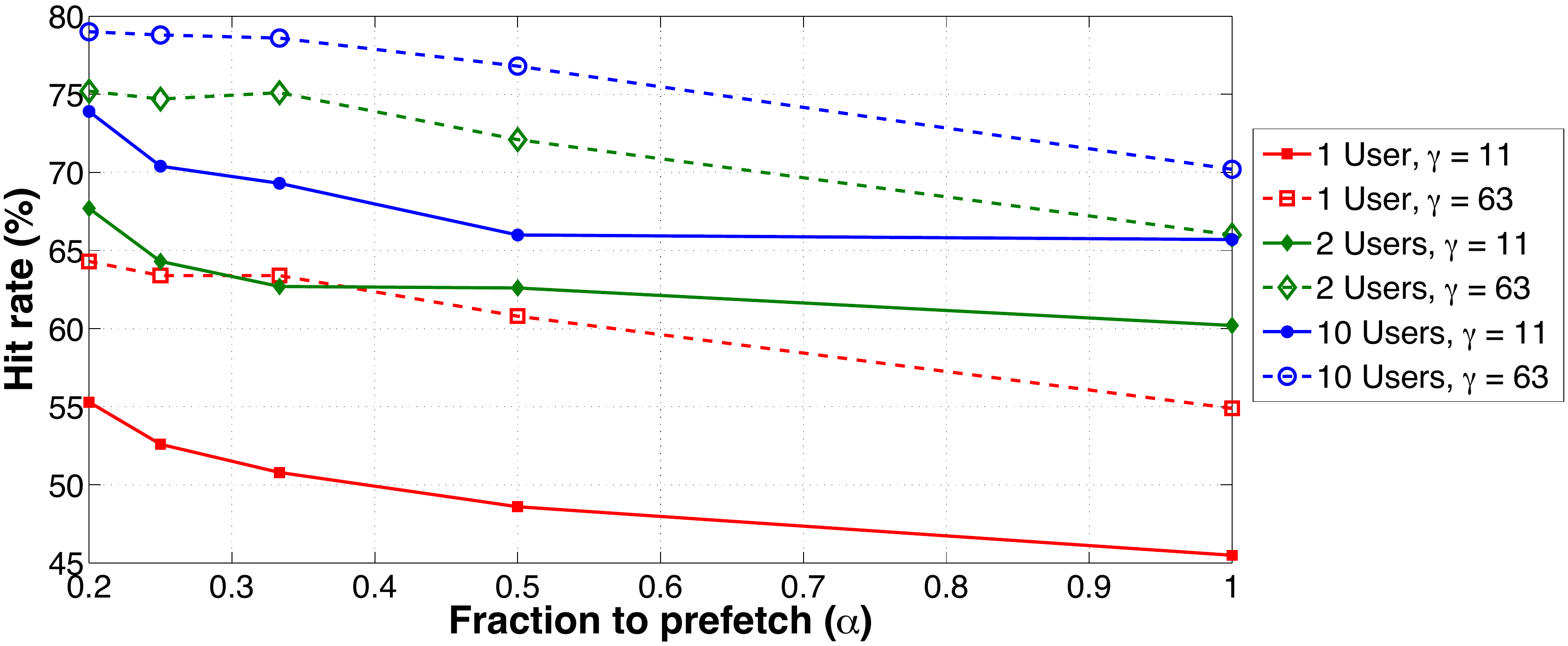}

\caption{Hit rate vs Fraction to pre-fetch $(\alpha)$ for a system with Number of videos $= 1000$, $P_{cont}  = 0.4$, Zipf parameter $(\beta) = 0.8$ and Cache size $= 200$. Hit-rates decrease with increasing values of $\alpha$ as the number of recommendations to pre-fetch reduce as $\alpha$ increases. }\label{fig:HitvsAlpha} 
\end{figure}

In the simulation results discussed so far, we pre-fetch complete videos. We now explore the possibility of pre-fetching only a fraction of the video and fetching the remaining part of the video only after the request is made. If there exists an $\alpha < 1$ such that while the user watches the first $\alpha$ fraction of the video, the remaining $(1-\alpha)$ fraction of the video can be pre-fetched, the CDN can provide uninterrupted service to the user without any startup delay by pre-fetching only the first $\alpha$ fraction of the video. 


Pre-fetching only a fraction of video reduces the bandwidth usage, thus reducing the overall cost of service as shown in Figures \ref{fig:CostvsAlpha12} and \ref{fig:CostvsAlpha64}. Since the bandwidth usage per pre-fetch is reduced, this allows the CDN to pre-fetch more recommendations at the same cost (Figure \ref{fig:RecosvsAlpha}) which leads to improved cache hit rates (Figure \ref{fig:HitvsAlpha}).

\section{Alternative Model}
\label{section:directed_models}

In Section \ref{section:our_model}, we used the Barabasi-Albert model to generate the recommendation graph $G(V,E)$. The Barabasi-Albert model generates an undirected graph. In the model described in \ref{section:our_model}, we replace each edge of this graph by two directed edges to get the recommendation relationships. As a result, if $v_i$ recommends $v_j$, then, $v_j$ also recommends $v_i$. However, the recommendation links in a VoD service may not always be bidirectional. In this section we explore a directed graph model that can be used to capture this property.

\subsection{Model Definition}
\label{subsection:model_dir}
Motivated by the fact that the degree distribution of the recommendation graph of VoD services follows the power law \cite{sweetyshubham2016}, instead of using the Barabasi-Albert model, we use a variation of the directed random graph model proposed in \cite{bollobas2003directed} for which the in-degree distribution follows the power law. The random graph is generated in an iterative manner by adding one node at the time. We start with an initial graph of $m$ nodes. Each new node is connected to the graph via $m$ edges. When a new node is introduced, edges are added to the graph in a sequential manner until there are $m$ edges involving the new node as follows:
\begin{enumerate}
\item[--] With probability $p_{out}$ a link from new node is created to an existing node $v$. Node $v$ is chosen randomly in proportion to the in-degree of $v$. 
\item[--] With probability $p_{in}$ a link from an existing node $v$ to the new node is created. Node $v$ is chosen randomly in proportion to the out-degree of $v$.  
\item[--] Otherwise, a link between 2 existing nodes $u$ and $v$ is created. The originating node (node $u$) and target node (node $v$) are chosen randomly in proportion to the in-degree and out-degree of $u$ and $v$ respectively.
\end{enumerate}
A formal description of the algorithm is given in Figure \ref{fig:Directed_graph_model}. 
\begin{figure}[!hb]
	\hrule
	\vspace{0.1in}
	\begin{algorithmic}[1]
		\STATE Initialize: Generate a connected graph of $m$ nodes ($v_1$, $v_2$, ..., $v_m$). Let $v=m+1$.
		\STATE Introduce a new node $n_v$ in the graph. Until the node gets connected with $m$ existing nodes in the graph, either via inlinks or outlinks, new links are sequentially added as follows:
		With probability $p_{out}$, a link from $n_v$ to $n_i$ for $i < v$ is added, the probability $p_i$ of choosing node $i$ is given by 
		$$p_i = \dfrac{K_i}{\sum_{j} K_j}, $$ 
		where $K_i$ is the current in-degree of node $n_i$. \\
		With probability $p_{in}$, a link from $n_i$ to $n_v$ is added, the probability $q_i$ of choosing node $i$ is given by	
		$$q_i = \dfrac{L_i}{\sum_{j} L_j}, $$ 
		where $L_i$ is the current out-degree of node $n_i$. \\
		With probability $1 - p_{out} - p_{in}$, a link from $n_i$ to $n_j$ is added for $i,j \leq m$, the probability $p_i$ of choosing node $n_i$, and $q_j$ of choosing node $j$ is given by
		 	$$p_i = \dfrac{K_i}{\sum_{j} K_j}, $$ 
		 	$$q_j = \dfrac{L_j}{\sum_{k} L_k}, $$ 
		\STATE $v = v+1$. If $v < n$, goto Step 2.
	\end{algorithmic}
	\vspace{0.1in}
	\hrule
	\caption{A directed random graph model which generates a random small-world directed graph with a degree distribution following the power law.}
	\label{fig:Directed_graph_model}
\end{figure}

Once the recommendation graph is generated, we assign transition probabilities to the various edges as discussed in Section \ref{section:our_model}. This completes the definition of our directed graph model.

\subsection{Properties}
Similar to the model proposed in Section \ref{section:our_model}, the alternative model also uses the empirically observed properties that the content popularity in absence of recommendation follows the Zipf's distribution and that chain count is between $1.3$ and $2.4$ to assign transition probabilities. Next, we verify if this model which uses a directed random graph model to generate the recommendation graph satisfies the rest of the empirical properties observed in section \ref{subsection:empirical_studies}. 
\subsubsection{Degree distribution}
\begin{figure}[h]
\centering
\includegraphics[width = 0.7\textwidth]{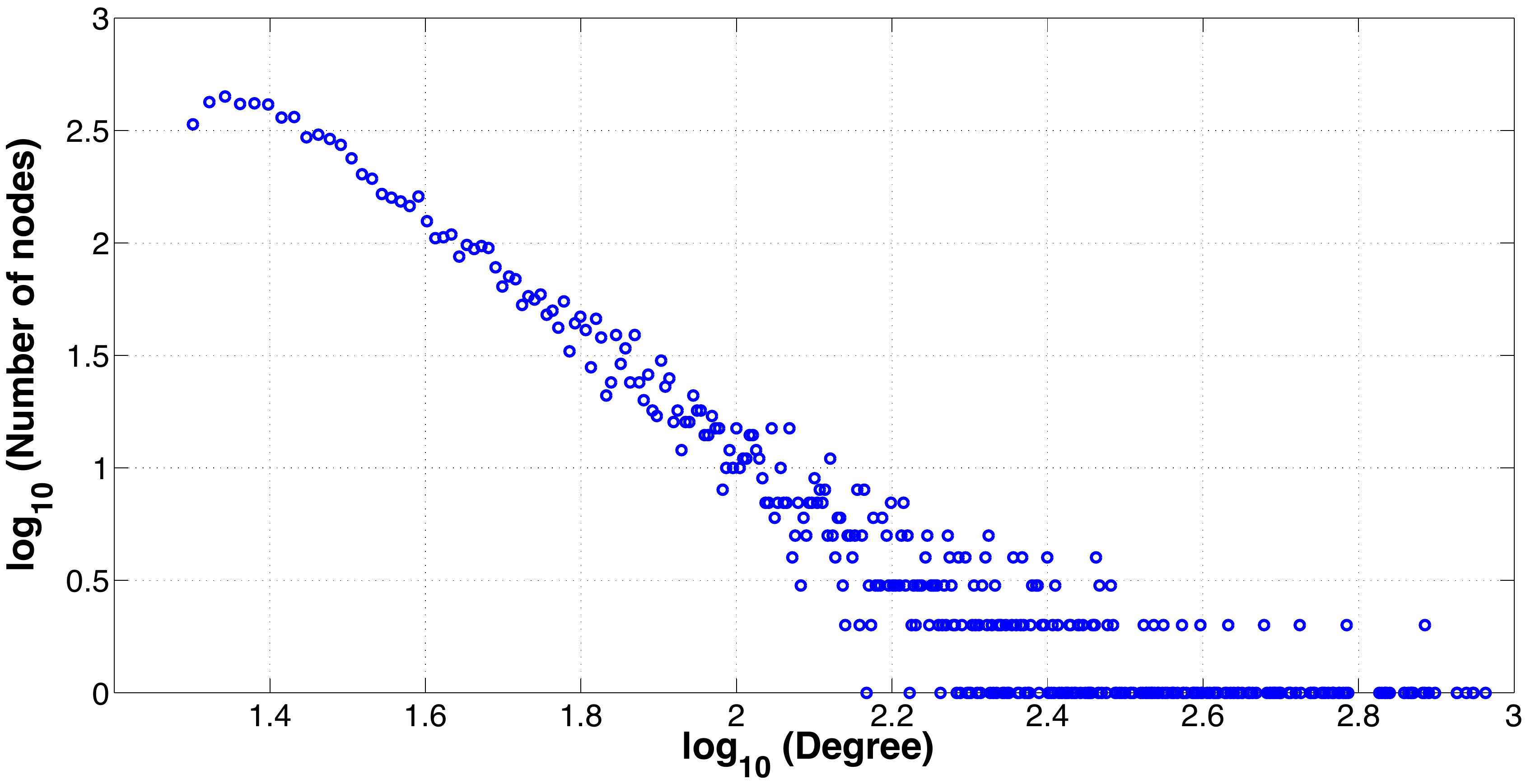}

\caption{In-degree distribution for the directed graph model of size 10,000 nodes with $p_{in} = 0.4$ and $p_{out} = 0.4$.}\label{fig:degree_distribution_krap_redner} 
\end{figure}

The graph $G(V,E)$ generated as described in Figure \ref{fig:Directed_graph_model} has a power law in-degree distribution, and the average number of out-links from a node is more than $mp_{out}$. Figure \ref{fig:degree_distribution_krap_redner} illustrates the in-degree distribution for a graph of $10,000$ nodes. The slope of the curve can be changed by changing the values of $p_{in}$ and $p_{out}$.

\subsubsection{Small World nature}
We evaluate the clustering coefficient and average path length for this graph with $2000$ nodes, $p_{in} = 0.4$ and $p_{out} = 0.4$. It is observed that this graph has a larger clustering and shorter average path lengths with respect to the Barabasi-Albert graph of the same size. We thus conclude that the construction in \ref{subsection:model_dir} generates a small world graph. 

\subsubsection{Content popularity profile}
\begin{figure}[h]
\centering
\includegraphics[width = 0.7\textwidth]{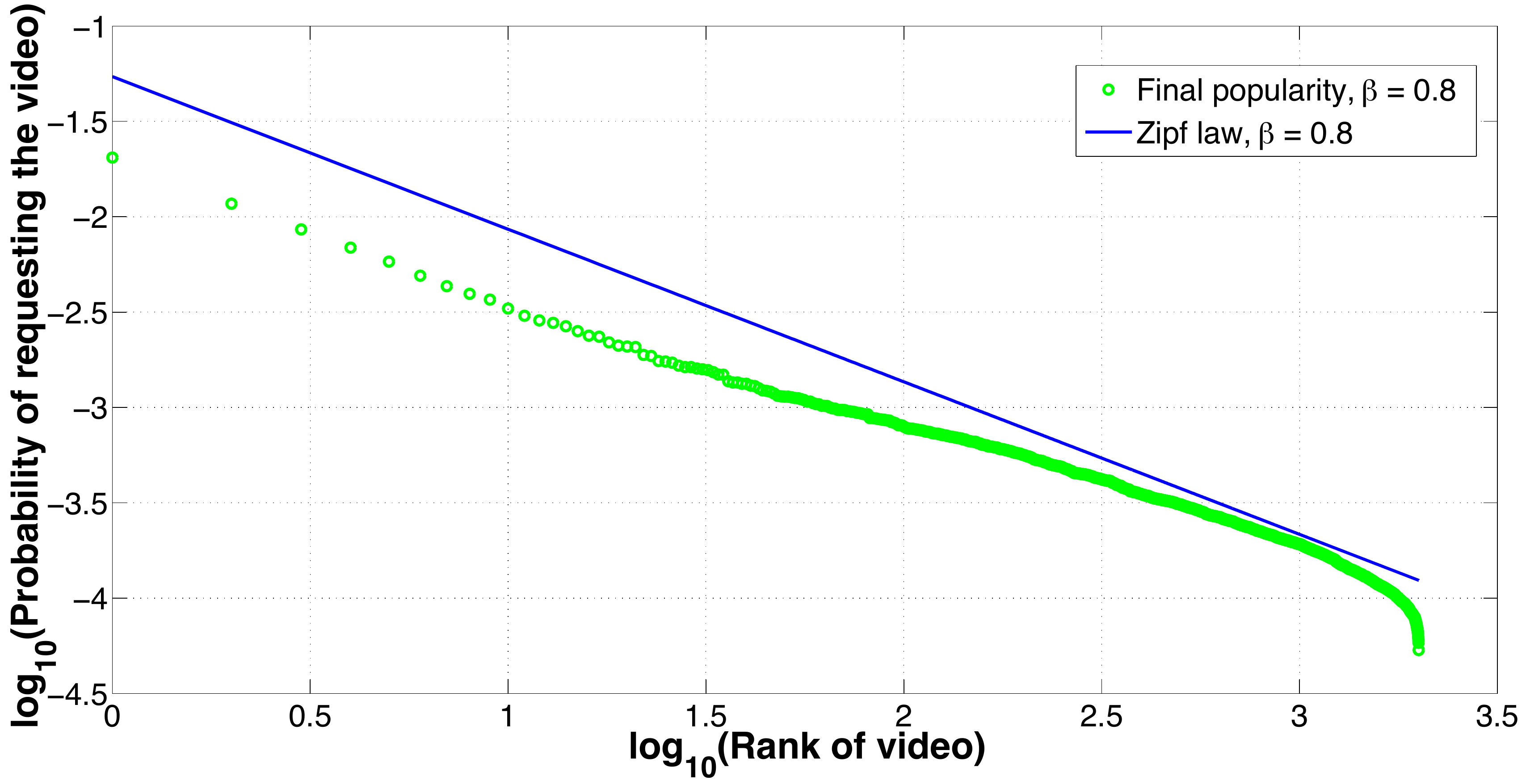}

\caption{Content popularity profile for the model with $m = 40$, $p_{out} = 0.4$, $p_{in} = 0.4$, Zipf parameter ($\beta$) $= 0.8$, number of videos $(n) = 2000$ and $\kappa = 0.8$}\label{fig:final_popularity_directed_model} 
\end{figure}

We generate the content popularity profile of our model by calculating the stationary distribution of the Markov Chain (as in Section \ref{section:our_model}). Figure \ref{fig:final_popularity_directed_model} shows the content popularity profile in our model. We see that, the content popularity profile follows the Zipf distribution for the popular videos and  decreases faster than as predicted by the Zipf distribution for the unpopular videos. Therefore, we conclude that content popularity profile for the alternative model is consistent with the observations in \ref{subsection:empirical_studies}. 

\subsubsection{Click through rate}
\begin{figure}[!hb]
\centering
\includegraphics[width = 0.7\textwidth]{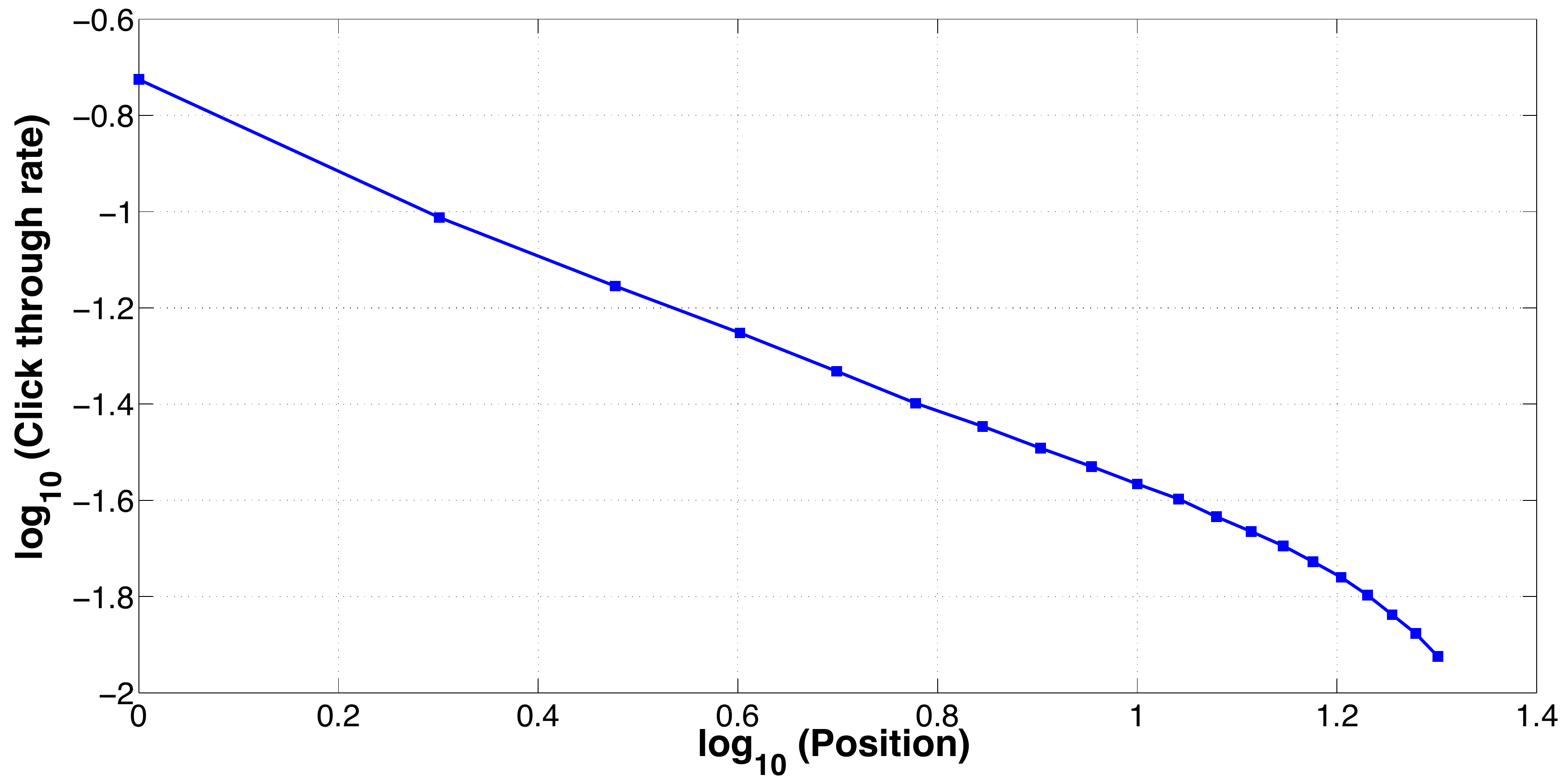}

\caption{CDF of CTR vs position in recommendation list for Number of videos ($n$) = 2000, $m = 40$, $p_{in} = 0.4$, $p_{out} = 0.4$, Zipf parameter $\beta = 0.8$, $\kappa = 0.8$, and $P_{cont} = 0.4$.}\label{fig:click_through_rate_directed_model} 
\end{figure}

As in section \ref{section:our_model}, we plot the median Click Through Rate (CTR) for the directed graph model in Figure \ref{fig:click_through_rate_directed_model}. We see that the median CTR can be approximated by the Zipf distribution. Our model is therefore consistent with the observations in \ref{subsection:empirical_studies}.

\subsubsection{Fraction of bi-directional links}
\begin{figure}[!hb]
\centering
\includegraphics[width = 0.7\textwidth]{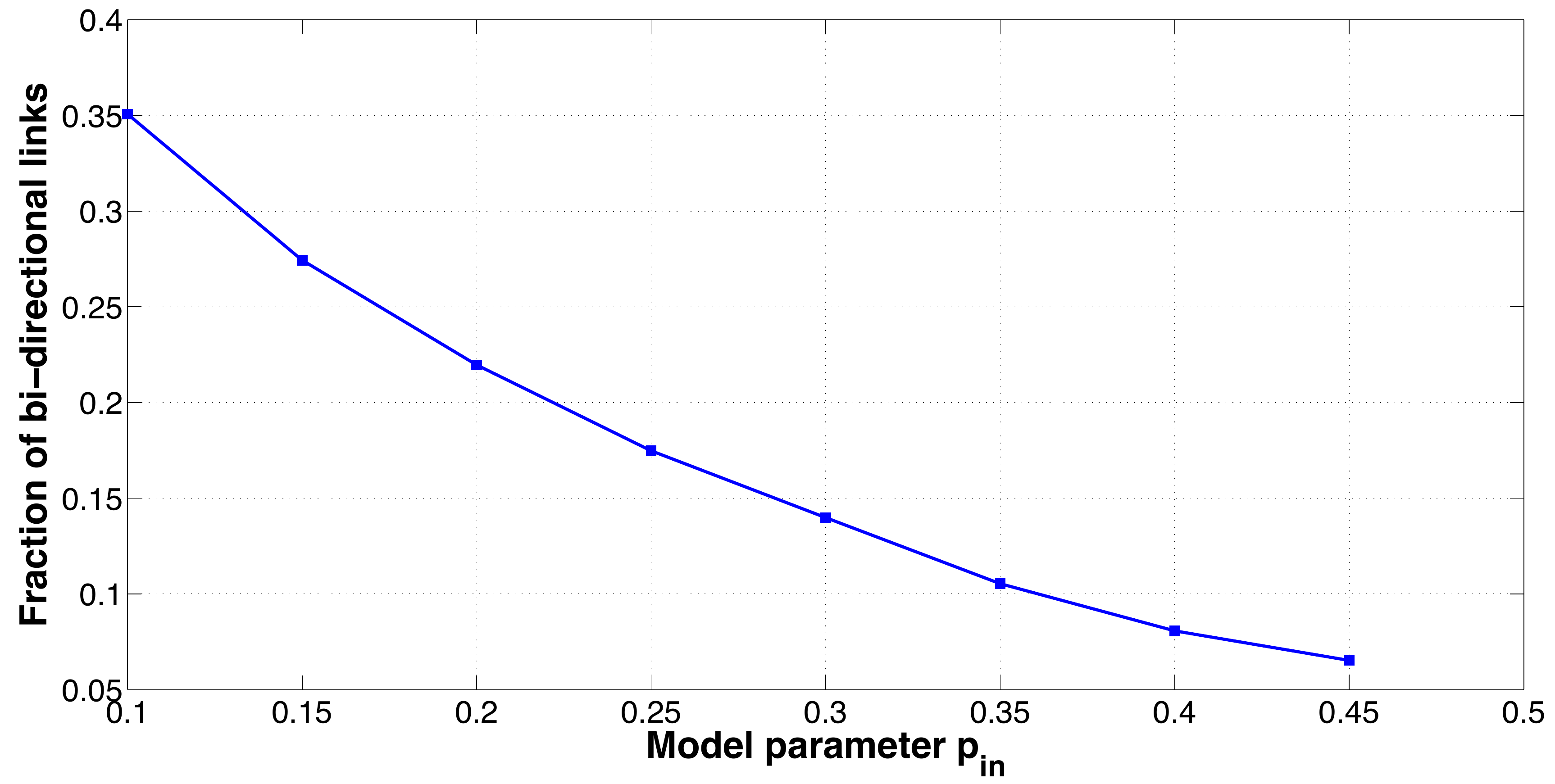}

\caption{Fraction of bi-directional links as a function of model parameter $p_{in}$ for a graph of size $2000$ nodes and $p_{out} = p_{in}.$}\label{fig:bi_directional_links} 
\end{figure}

\begin{figure}[!hb]
\centering
\includegraphics[width = 0.7\textwidth]{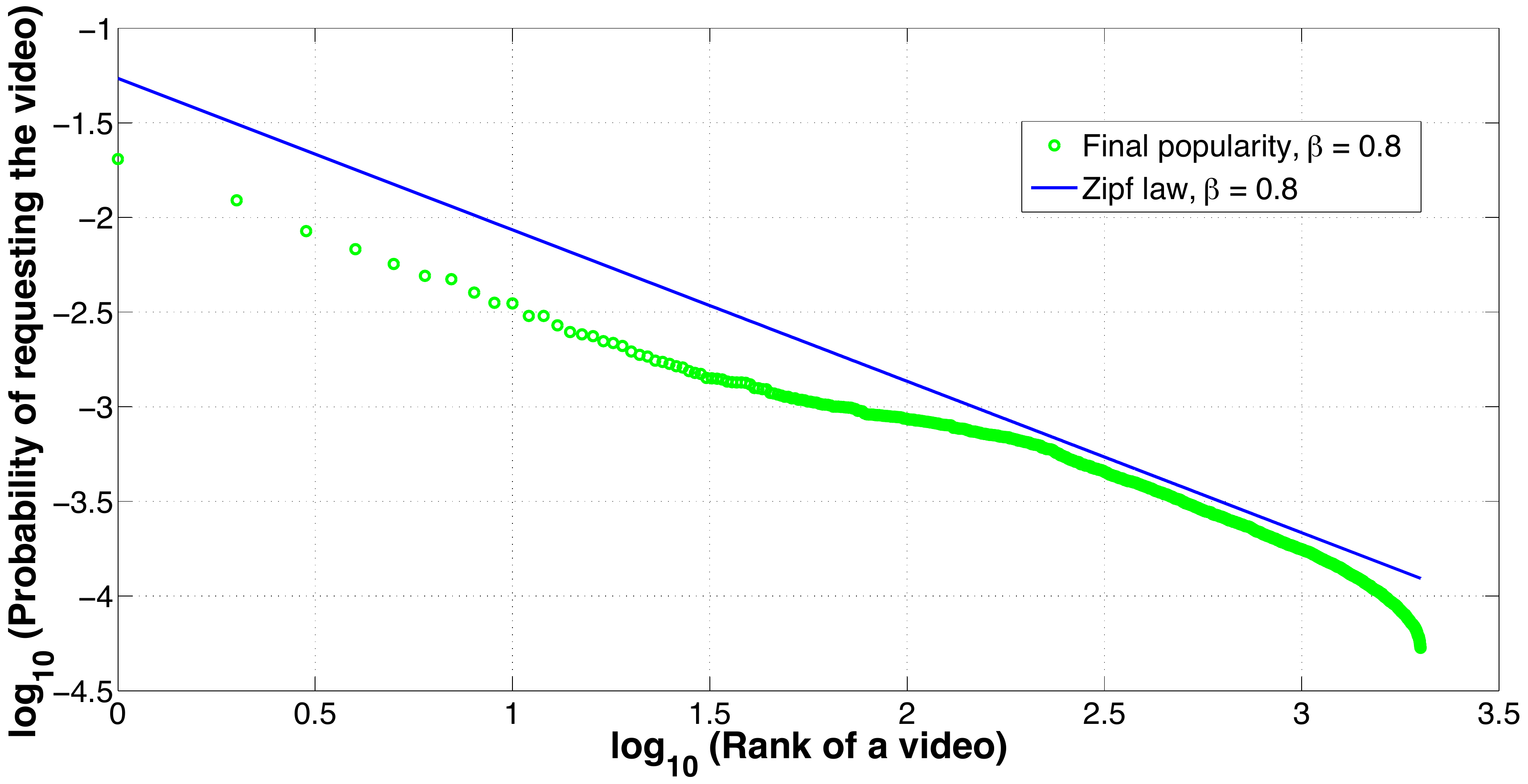}

\caption{Content popularity for the model with $m = 40$, $p_{out} = 0.2$, $p_{in} = 0.2$, Zipf parameter $(\beta) = 0.8$, number of videos $(n) = 2000$ and $\kappa = 0.8$. The content popularity gets distorted from the Zipf law for small values of $p_{in}$.}\label{fig:distorted_final} 
\end{figure}

From a small experiment on a sub-graph of the YouTube recommendation graph, we observed that about $~30\%$ of the recommendation links are bi-directional. Figure \ref{fig:bi_directional_links} shows the relationship between the model parameter $p_{in}$ and the fraction of bi-directional links in the graph. We conclude that the value of $p_{in}$ can be tuned to obtain the desired fraction of bi-directional links. This flexibility does not exist in the model proposed in Section \ref{section:our_model} and therefore, is a key point of difference between the model proposed in Section \ref{section:our_model} and the alternative model discussed in this section.
An issue with small values of $p_{in}$ is that it distorts the final popularity from the scale-free nature as shown in Figure \ref{fig:distorted_final}.

\subsection{Caching Simulations}
In this section, we study the performance of the LRU and PreFetch policy when requests arrive according to the model described in \ref{subsection:model_dir}. We use the same CDN and simulation setting as described in Sections \ref{section:CDN} and \ref{section:simulate} respectively. 

\begin{figure}[!hb]
\centering
\includegraphics[width = 0.7\textwidth]{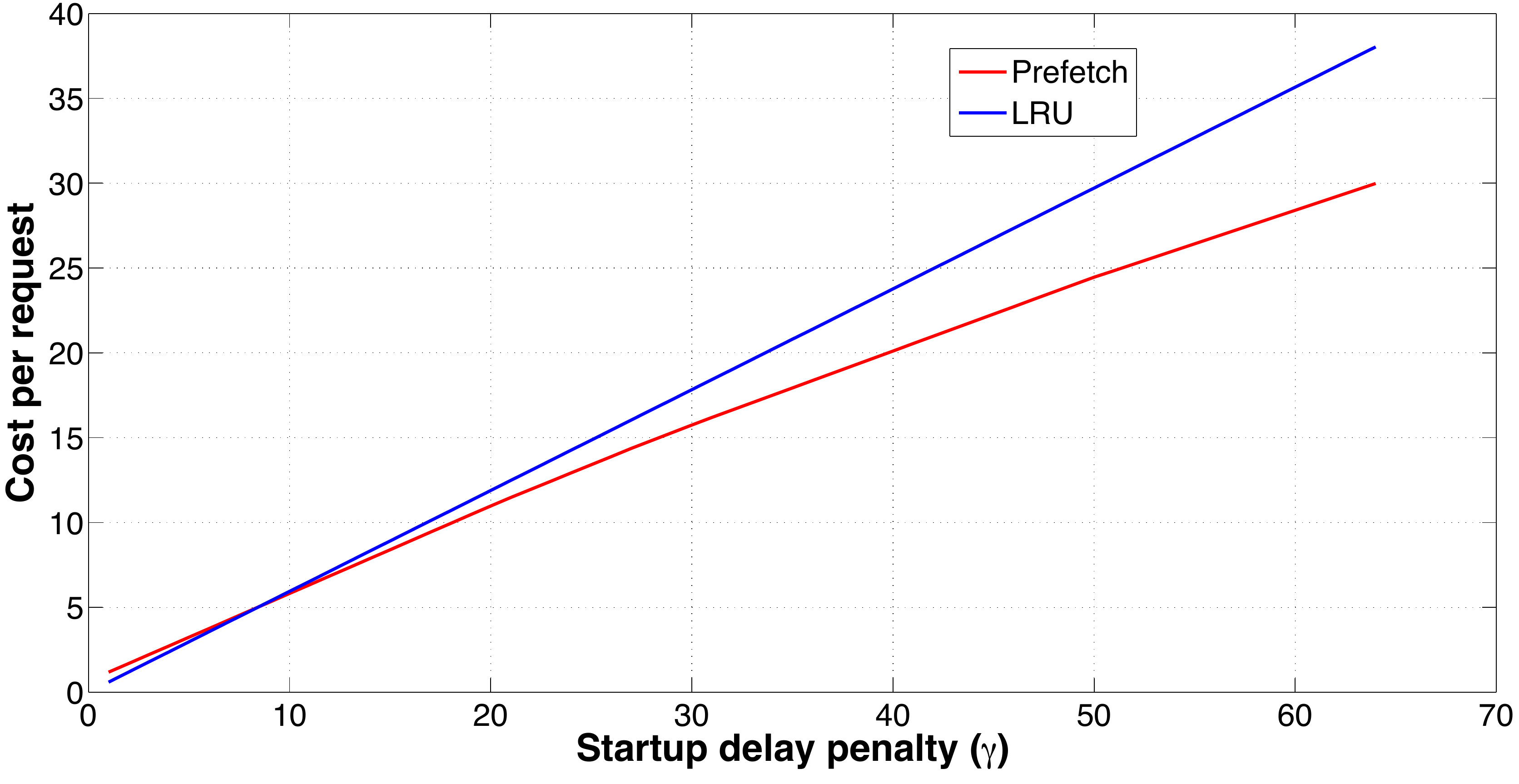}

\caption{Cost vs Start-up delay penalty ($\gamma$) for a system with number of videos $= 1000$, $m = 40$, Zipf parameter $(\beta) = 0.8$, cache size $= 200$, $P_{cont} = 0.4$, $p_{in} = 0.4$, $p_{out} = 0.4$ and $1$ user. As $\gamma$ increases, PreFetch outperforms LRU policy.  }\label{fig:directed_cost_v_gamma} 
\end{figure}

\begin{figure}[h]
\centering
\includegraphics[width = 0.7\textwidth]{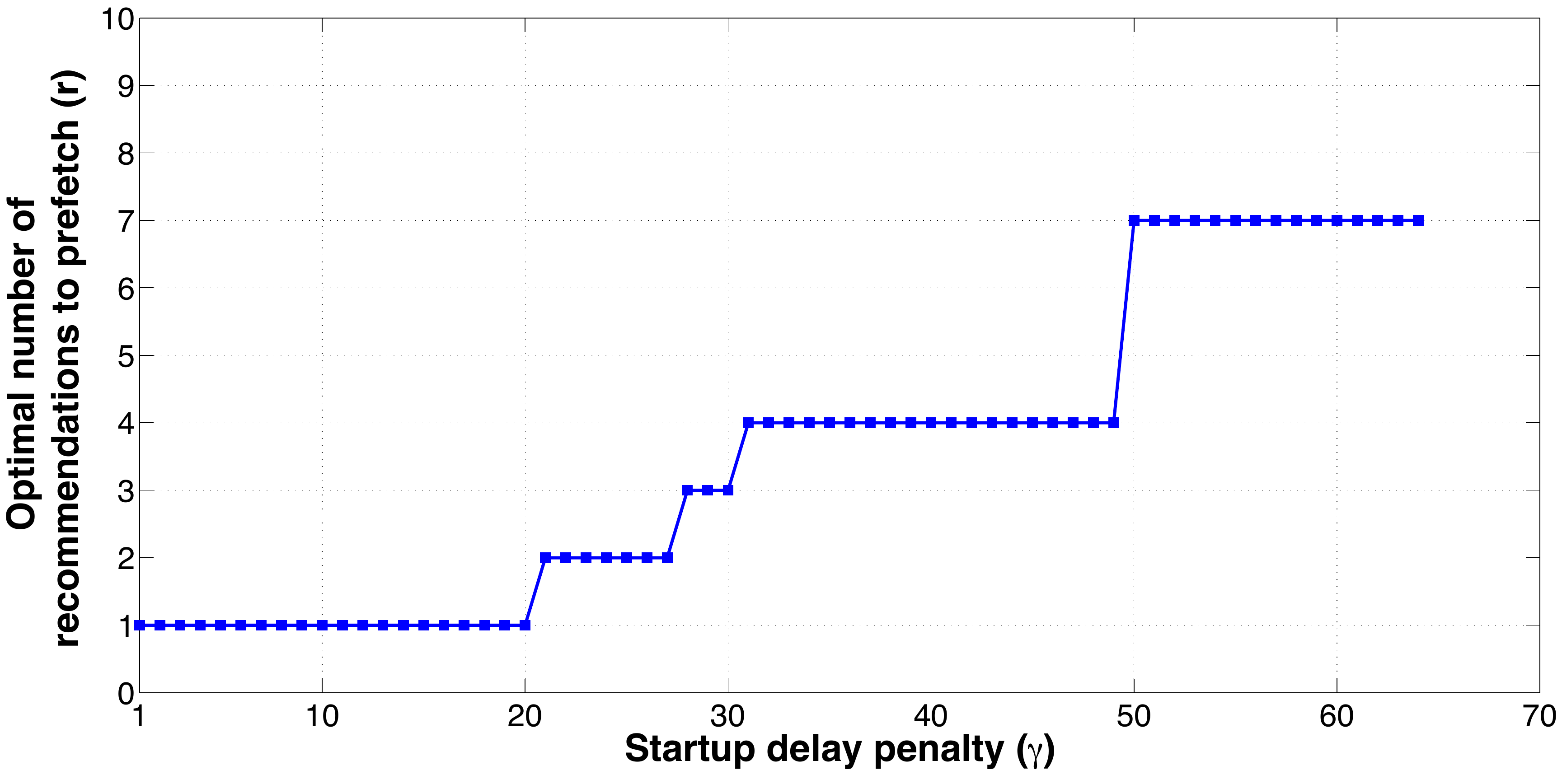}

\caption{The optimal number of recommendations to pre-fetch ($r$) vs Start-up delay penalty ($\gamma$) for a system with number of videos $= 1000$, $m = 40$, Zipf parameter$ (\beta) = 0.8$, cache size $= 200$, $P_{cont} = 0.4$, $p_{in} = 0.4$, $p_{out} = 0.4$ and $1$ user. The optimal number of recommendations to pre-fetch ($r$) increases with start-up delay penalty $\gamma$. }\label{fig:directed_r_v_gamma} 
\end{figure}

In Figure \ref{fig:directed_cost_v_gamma}, we compare the performance of PreFetch policy and the LRU policy as a function of the Startup delay penalty ($\gamma$). In Figure \ref{fig:directed_cost_v_gamma}, we use the empirically optimized value of $r$ (number of recommendations to prefetch) which leads to the lowest cost of service. The optimal number of recommendations to pre-fetch ($r$) increases with increase in Start-up delay penalty ($\gamma$) as shown in figure \ref{fig:directed_r_v_gamma}. Note that these plots exhibit qualitatively similar behavior as in Figures \ref{fig:CostvsW} and \ref{fig:RecosvsW}. 

The dependence on other parameters like number of users using a local server ($u$), size of cache, fraction of videos pre-fetched ($\alpha$) etc. is qualitatively similar to the results in Section \ref{section:simulate}.

\section{Conclusions}
\label{section:conclusions}

In this work, we propose a Markovian model for request arrivals in VoD services with recommendation engines which captures the time-correlation in user requests and is consistent with empirically observed properties. 

Low start-up delay is a key QoS requirement of users of VoD services. In addition, minimizing the bandwidth consumption of the network is key to reduce the cost of service. Given the trade-off between these two goals, we show that the time-correlation in user requests can be used to design caching policies which outperform popular policies like LRU which do not exploit this time-correlation. More specifically, we show that our caching policy $\emph{PreFetch}$ which employs recommendation based pre-fetching outperforms the LRU policy in terms of the joint cost of start-up delay and bandwidth consumption when the relative is cost of start-up delay is high. 
  
\bibliographystyle{splncs_srt}
\bibliography{myref}

\end{document}